\documentclass[aps,twocolumn,prd,superscriptaddress]{revtex4-2}

\usepackage[utf8]{inputenc}

\usepackage{mathtools}
\usepackage{amsfonts}
\usepackage{mathrsfs}
\usepackage{bbm}
\usepackage{physics}
\usepackage{slashed}
\usepackage{tensor}

\usepackage{graphicx}
\usepackage{color, float}
\usepackage{array}
\usepackage[abs]{overpic}

\usepackage{placeins}

\usepackage{makecell}
\usepackage{subcaption}

\usepackage{xspace}
\usepackage{siunitx}
\usepackage{xfrac}
\usepackage{hyperref}
\usepackage[nameinlink]{cleveref}
\usepackage{appendix}

\usepackage{xifthen}
\usepackage{xcolor}
\hypersetup{
	colorlinks,
	linkcolor={red!75!black},
	citecolor={blue!75!black},
	urlcolor={blue!75!black}
}

\usepackage{booktabs}
\usepackage{multirow}

\newcolumntype{C}{>{$}c<{$}}
\AtBeginDocument{
	\heavyrulewidth=.08em
	\lightrulewidth=.05em
	\cmidrulewidth=.03em
	\belowrulesep=.65ex
	\belowbottomsep=0pt
	\aboverulesep=.4ex
	\abovetopsep=0pt
	\cmidrulesep=\doublerulesep
	\cmidrulekern=.5em
	\defaultaddspace=.5em
}

\graphicspath{{./../figures/},{./figures/}}

\newcommand{\gettitle}{Tunneling with physics-informed RG flows in the anharmonic oscillator}

\newcommand{\getCataniaAffiliation}{\affiliation{INAF, Osservatorio Astrofisico di Catania, via S. Sofia 78, 95123 Catania, Italy}}
\newcommand{\getCataniaSAffiliation}{\affiliation{INFN, Sezione di Catania, via S. Sofia 64, 95123 Catania, Italy}}
\newcommand{\getZuerichAffiliation}{\affiliation{Institut f{\"u}r Theoretische Physik, ETH Z{\"u}rich, Wolfgang-Pauli-Str. 27, 8093 Z{\"u}rich, Switzerland}}
\newcommand{\getHeidelbergAffiliation}{\affiliation{Institut f{\"u}r Theoretische Physik, Universit{\"a}t Heidelberg, Philosophenweg 16, 69120 Heidelberg, Germany}}
\newcommand{\getEMMIAffiliation}{\affiliation{ExtreMe Matter Institute EMMI, GSI, Planckstr. 1, 64291 Darmstadt, Germany}}

\hypersetup{
	pdftitle={\gettitle},
	pdfauthor={Bonanno, Ihssen, Pawlowski},
	pdfkeywords={anharmonic Oscillator}
	{physics-informed RG flows} {renomalisation group} ,
	bookmarksopen=true,
	bookmarksopenlevel=2,
	bookmarksnumbered=true
}

\begin{document}
	
\title{\gettitle}
	\author{Alfio Bonanno}\getCataniaAffiliation\getCataniaSAffiliation
	\author{Friederike Ihssen}
	\thanks{ihssen@thphys.uni-heidelberg.de}
	\getHeidelbergAffiliation\getZuerichAffiliation
	\author{Jan M. Pawlowski}\getHeidelbergAffiliation\getEMMIAffiliation
\begin{abstract}

We solve the anharmonic oscillator with physics-informed renormalisation group (PIRG) flows, with an emphasis on the weak coupling regime with its instanton-dominated tunnelling processes. We show that the instanton physics behind the exponential decay of the energy gap is already covered in the first order of the derivative expansion of the PIRG. The crucial new ingredients in the present analysis are the use of the ground state expansion within PIRG flows, as well as precision numerics based on Galerkin methods. 
Our result $a_{\mathrm{inst}} = 1.910(2)$ for the decay constant is in quantitative agreement with the analytic one, $a_{\mathrm{inst}} = 1.886$ with a deviation of $1\%$. This illustrates very impressively the capacity of the PIRG for fully capturing non-perturbative physics already in relatively simple approximations. 

\end{abstract}

	\maketitle

\section{Introduction}
\label{sec:Introduction}

In the past five decades, the functional renormalisation group (fRG) approach has led to an impressive plethora of non-perturbative results in statistical mechanics and quantum field theory, see \cite{Dupuis:2020fhh} for a recent compilation of results in various areas. As an exact equation it accommodates topological effects, and its capacity for fully non-perturbative computations has been impressively demonstrated in many areas of physics. Still, it is left to assess, which approximations are well-suited for capturing phenomena that are driven by topological configurations or defects.

An optimal test case for such an analysis is provided by the anharmonic oscillator with its instanton-dominated regime at small couplings $\lambda_\varphi$: Firstly, it can be solved numerically with Hamiltonian methods for all couplings. Secondly, in the weak coupling regime it has an asymptotic analytic solution obtained in a saddle-point expansion about the topological instanton solutions. Finally, the result for the energy gap $\Delta E$ between the ground state and the first excited state has a very characteristic non-perturbative dependence on the anharmonicity $\lambda_\varphi$ with 
\begin{align} 
\Delta E(\lambda_\varphi \to 0) \propto \frac{1}{\sqrt{\lambda_\varphi}}\,  e^{-a_{\mathrm{inst}}  /\lambda_\varphi }\,.
\label{eq:DefaInst}
\end{align} 
In \labelcref{eq:DefaInst} we have measured the coupling $\lambda_\varphi$ in units of the mass. The analytic result for the coefficient in the exponent is $a_{\mathrm{inst}} = 4\, \sqrt{2}/3\approx  1.886$, see \labelcref{eq:DeltaETop}. The exponential damping can be considered the smoking gun for the topological scaling induced by the tunnelling process: this behaviour with an essential singularity for $\lambda_\varphi \to 0$ cannot be obtained within perturbation theory and its signatures are well separated from the perturbative polynomial result. In combination this offers a complete insight to the system, and in particular also an analytic understanding of the underlying tunnelling physics. Accordingly, the anharmonic oscillator can be used as a perfect model to dissect the convergence of the fRG approach, applied to topological phenomena. Specifically, such an analysis answers the question, which approximation suffices to fully incorporate topological configurations. 

For these reasons the anharmonic oscillator has been studied with the fRG for the past two decades, initiated by the work in \cite{Zappala:2001nv}, for further works see e.g.~\cite{Aoki:2002ozs, Weyrauch:2006aj, Kovacs:2014mia, Baldazzi:2021guw, Bonanno:2022edf}. In these works the derivative expansion has been used, either in leading order (local potential approximation (LPA)) or in its first order with a field-dependent wave function. While the lowest order approximation fails to accommodate the instanton regime, the first order results point towards a sizeable improvement. Still, none of these computations could confirm the exponential decay: Firstly, the computations do not include the small coupling regime covered by the two-loop or one-loop saddle point expansion. Moreover, already for larger couplings outside the exponential regime, sizeable deviations from the correct scaling with the coupling have been reported. 

In the current work we follow up on these analyses and study the anharmonic oscillator in the first order of the derivative expansion. The analysis here is based on three pivotal new ingredients: 
\begin{itemize} 
\item[(i)] The use of physics-informed RG (PIRG) flows introduced in \cite{Ihssen:2024ihp} within an expansion about the ground state of the theory \cite{Ihssen:2023nqd}, see \Cref{sec:flowingField}. This framework is able to optimise the convergence of any expansion scheme and is applied to the derivative expansion. 
\item[(ii)] The use of powerful numerical techniques, see e.g.~\cite{Ihssen:2022xkr, Ihssen:2023qaq, Sattler:2024ozv}. Further details can be found in \Cref{app:Numerics}. 
\item[(iii)] The use of an alternative observable for the detection of the exponential decay: instead of only resolving the energy gap $\Delta E$, we are computing the coupling-dependence of the size of the exponentially flat field regime about vanishing fields. The two are directly related since the energy gap is proportional to the curvature of the potential at vanishing field value and hence signals an exponentially flat regime.
\end{itemize} 
The use of \textit{(i)} and \textit{(ii)} allows us to extend previous analyses to far smaller couplings, also reaching for the instanton-dominated small coupling regime within the first order of the derivative expansion. Then, \textit{(iii)} makes use of an optimal observable that does not necessitate the evaluation of the potential in the exponentially flat regime itself. In combination, these three novel ingredients allow us to resolve the instanton-dominated regime of the anharmonic oscillator. In particular we confirm the persistence of the exponential flattening of the energy gap in the current PIRG approach, see \Cref{sec:InstantonRegime}. This clearly signifies that the fRG accommodates the effects of topological configurations already in relatively simple approximations. Moreover, we can convert the coupling dependence of the scaling of the exponentially flat regime to that of the energy gap. This leads to the result $a_{\mathrm{inst}} = 1.910(2)$, see \labelcref{eq:ainstPIRG} with a $1\%$ deviation from the analytic result $a_{\mathrm{inst}} \approx 1.886$ mentioned above.

\section{The anharmonic oscillator and the derivative expansion}
\label{sec:Anharmonic+Derivative}

The quantum mechanical anharmonic oscillator can be expressed as a 0+1 dimensional Euclidean quantum field theory with the classical action 
\begin{align}
	S[\hat\varphi] = \int d \tau \left\{ \frac{1}{2} \left(\partial_\mu \hat\varphi\right)^2 +\frac{m_\varphi^2}{2} \hat\varphi^2 + \frac{\lambda_\varphi |m_\varphi|^3 }{8} \, \hat\varphi^4\right\} \,,
	\label{eq:ClassicalAction1-ON}
\end{align}
with the anharmonic self-interaction part and imaginary time $\tau$. For later convenience we have expressed \labelcref{eq:ClassicalAction1-ON} in the dimensionless coupling $\lambda_\varphi$, measured in units of the mass scale $|m_\varphi|$, by scaling out the momentum dimension with $|m_\varphi|^3$. From now on we simply express all scales in units of the mass and \labelcref{eq:ClassicalAction1-ON} turns into 
\begin{align}
	S[\hat\varphi] = \int d \tau \left\{ \frac{1}{2} \left(\partial_\mu \hat\varphi\right)^2 -\frac12  \hat\varphi^2 + \frac{\lambda_\varphi }{8} \, \hat\varphi^4\right\} \,,
	\label{eq:ClassicalAction1-ONDimless}
\end{align}
where the minus sign in front of the mass term $1/2  \hat\varphi^2$ indicates the negative mass squared. The field $\hat\varphi$ in \labelcref{eq:ClassicalAction1-ONDimless} is the microscopic quantum field or operator with the mean field $\varphi = \langle \hat\varphi\rangle$ . In the anharmonic oscillator it is nothing but the time-dependent position variable $x(\tau)=\varphi(\tau)$. 

For double-well potentials $V_\textrm{cl}(\varphi)$ with $m_\varphi^2<0$, the classical potential has two minima at $\pm \varphi_0$ with 
\begin{align} 
	\varphi_0 = \sqrt{\frac{2}{\lambda_\varphi}}\,. 
\label{eq:varphi0Classical}
\end{align} 
We emphasise that this minimal field $x_0=\varphi_0$ has a direct physical meaning in our quantum mechanical example: it indicates the spatial location of the minimum of the potential. This has to be contrasted with a higher dimensional quantum field theory, where the field amplitude is not a direct measurable quantity such as a location. 

\begin{figure}
	\includegraphics[width=.98\columnwidth]{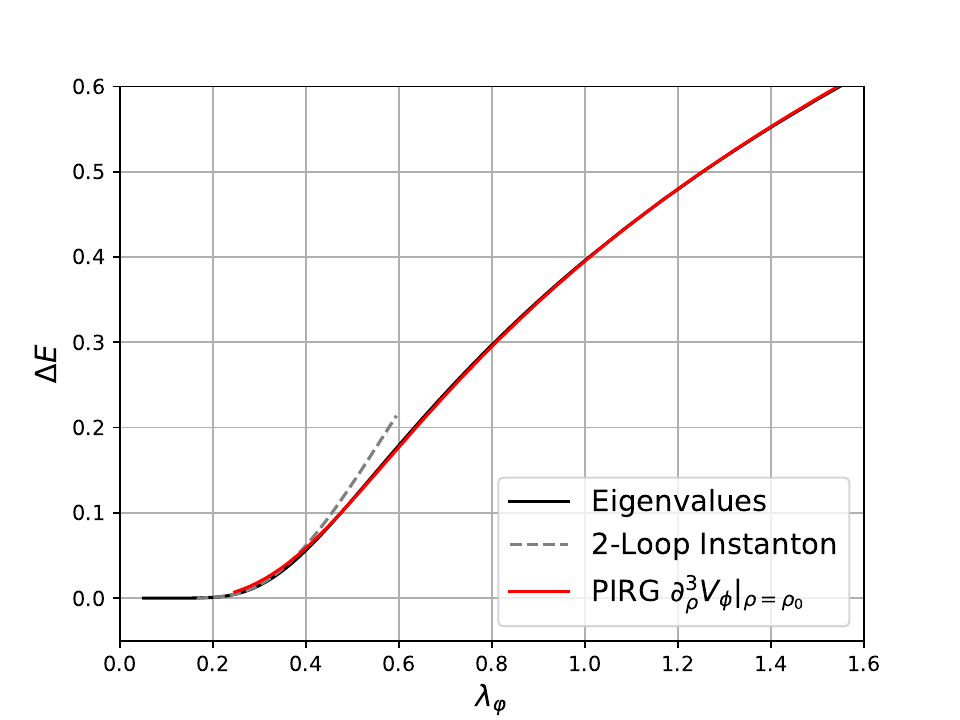}
	\caption{Energy gap $\Delta E$ between the ground state and the first excited state: \textit{Eigenvalues:} Numerical evaluation of the energy eigenvalues of the Schrödinger equation (black line), \textit{Instanton:} two-loop saddle point expansion in the dilute gas expansion \labelcref{eq:TwoLoopInst} (dashed grey line), \textit{PIRG:} Basic approximation to the first order derivative expansion of physics-informed flows in the ground state expansion (red straight line), see \Cref{app:EnergyGapPIRG}. The novel observable sensitive to the topological scaling is introduced in \Cref{sec:ExponentialPIRG}, see \Cref{fig:pltos,fig:arho0aDeltaE} . \hspace*{\fill}}
	\label{fig:Instanton+Schroedinger+PIRG}
\end{figure}
In quantum mechanics spontaneous symmetry breaking is absent as tunnelling between the two minima requires a finite energy and the ground state is a superposition of the states centred at $\varphi_0$. For small anharmonicities $\lambda_\varphi \to 0$, these minima are well separated. In this limit the tunnelling phenomenon even dominates the fluctuation physics. 

This is clearly visible in the energy gap $\Delta E$ between the ground state and the first excited state, which carries a clear signature for the topological tunnelling effects. In the above-mentioned limit of small anharmonicities it can be computed analytically using a saddle point expansion about the topological tunnelling solution, the quantum-mechanical instanton. The dimensionless energy gap $\Delta E$ within the two-loop dilute instanton gas approximation reads \cite{Schafer:1996wv}, 
\begin{subequations} 
	\label{eq:DeltaETop}
\begin{align}
\Delta E_\textrm{inst}= 8 \ 2^{1/4} \sqrt{\frac{1}{\pi\, \lambda_\varphi}} \, e^{-S_0 -\frac{71}{72} \frac{1}{S_0}} , \quad  S_0 = \frac{4}{3} \frac{\sqrt{2}}{\lambda_\varphi} \,,
	\label{eq:TwoLoopInst}
\end{align}
with the instanton action $S_0$ and the exponent 
\begin{align} 
	S_0 + \frac{71}{72} \frac{1}{S_0} =   \frac{a_\textrm{inst}}{\lambda_\varphi}+ \frac{71}{96}\lambda_\varphi \approx 
	\frac{1.886}{\lambda_\varphi} + 0.523\, \lambda_\varphi \,,
\label{eq:ExponentDeltaE} 
\end{align} 
\end{subequations} 
measured in units of $|m_\varphi|$.
Importantly, the expression \labelcref{eq:DeltaETop} displays a characteristic exponential suppression, which cannot be obtained in perturbation theory about the trivial vacuum: all expansion coefficients of a Taylor expansion about $\lambda_\varphi=0$ vanish identically. In turn, for large anharmonicities perturbation theory works very well, and 	\labelcref{eq:DeltaETop} fails. 

Both asymptotic regimes and the intermediate transition regime can be obtained accurately by evaluating the energy eigenvalues of the Schrödinger equation numerically. In the present work we use a basis consisting of the 50 lowest eigenstates to compute a benchmark for the energy gap, see \Cref{fig:Instanton+Schroedinger+PIRG}. From this plot it is also evident that we have to go to anharmonicities smaller than $0.4$, to be safely in the instanton-dominated topological regime, 
\begin{align}
	 \lambda_\varphi\lesssim \lambda_\textrm{inst} \,,\qquad \textrm{with} \qquad  \lambda_\textrm{inst} \approx 0.4\,, 
	\label{eq:lambda_tun}
\end{align}
see the analysis in \Cref{sec:MappingOutInstanton}. We also show a basic approximation to $\Delta E$ obtained in the present work, see \Cref{app:EnergyGapPIRG}. As is explained there, this basic approximation captures the perturbative regime quantitatively, while it only shows qualitative properties in the instanton-dominated topological regime. The novel observable, designed for the computation of $a_\textrm{inst}$ in \labelcref{eq:DeltaETop}, is introduced in \Cref{sec:ExponentialPIRG}, see \Cref{fig:pltos,fig:arho0aDeltaE}. This finalises our discussion of the benchmarks, both the numerical solution and the topological analytic one in the instanton dominated regime. 

The computation in the present work is done with the functional renormalisation group equation for the effective action, the Legendre transform of the logarithm of the generating functional. As mentioned in the introduction, we use the novel physics-informed setup introduced in \cite{Ihssen:2024ihp} and explained in \Cref{sec:PIRGs}. Moreover, we use the derivative expansion for the effective action. This is an expansion of the effective action in powers of derivatives of the field, that is in terms of $p^2/m_\textrm{gap}^2$ with a given mass gap $m_\textrm{gap}$. 
The fRG approach is very well adapted to such an expansion scheme as the infrared cutoff itself is an addition to the physical mass gap of the theory, roughly speaking $m_\textrm{gap}^2 \to m_\textrm{gap}^2+k^2$. In this approximation, the effective action is given by 
\begin{align}
	\Gamma_\varphi[\varphi] = \int d\tau \left[ \frac12 Z_\varphi(\varphi) (\partial_\mu \varphi)^2 +V_\varphi(\varphi)+\cdots\right]\,. 
	\label{eq:Gamma1stOrder}
\end{align}
\Cref{eq:Gamma1stOrder} displays the first order of the derivative expansion: the full effective potential $V_\varphi(\varphi)$ and the classical kinetic term constitute the zeroth order term, also called the local potential approximation (LPA). The first order term is proportional to $Z_\varphi(\varphi)-1$. The dots comprise all higher order terms starting with $W(\varphi) (\partial_\mu^2 \varphi) (\partial_\mu^2\varphi)$. Finally, the subscript ${}_\varphi$ indicates that the Legendre transform is taken with respect to the fundamental field. 

The energy gap between the ground state and first exited state of the anharmonic oscillator is directly given by the mass of the scalar propagator. In terms of the effective action \labelcref{eq:Gamma1stOrder}, it is given by the RG-invariant curvature of the effective potential, evaluated at the equations of motion $\varphi=0$, 
\begin{align}
	\Delta E = \sqrt{\frac{V_\varphi''(0)}{Z_\varphi(0)}} \,.
	\label{eq:DeltaE}
\end{align}
\Cref{eq:DeltaE} has been computed in LPA and in the first order derivative expansion for the effective action $\Gamma_\varphi[\varphi]$. However, while for $\lambda_\varphi$ far outside the topological regime \labelcref{eq:lambda_tun} and for the whole coupling regime with $m_\varphi^2 >0$, the first order results agree very well. Close to the topological regime the results show sizeable deviations. Moreover, the standard setup does not allow us to dive deep into the topological regime but only to graze it. 

This concludes our brief overview of the anharmonic oscillator.

\section{Physics-informed RG flows and the ground state expansion}
\label{sec:flowingField}

In the following we set up a PIRG approach which overcomes these limitations and, with the choice of an appropriate observable, allows us to quantitatively determine the prefactor in the exponent of \labelcref{eq:TwoLoopInst}. To this aim we resolve the exponentially flat regime between the minima $\pm \varphi_0$ in the effective potential of the anharmonic oscillator, which arises in the regime dominated by instanton-induced tunnelling processes. Evidently, the combination of an exponentially flat regime for small fields and a polynomial regime for larger fields destabilises Taylor expansions in the field. About $\varphi=0$, it does not capture the physics at all. Moreover, even within computations of the full effective potential using advanced numerical methods this combination of different regimes leads to instabilities if going beyond the lowest order of the derivative expansion. For explicit examples of the potential for different couplings see \Cref{fig:Potentials} in \Cref{sec:Numerics}. 

This calls for an expansion scheme with optimised convergence, and in the present work we use recent advances in physics-informed RG flows (PIRG flows) \cite{Ihssen:2024ihp}, or more precisely the ground state expansion in PIRGs \cite{Ihssen:2023nqd}. In this approach the derivative expansion is done about the full propagator of the theory which is cast in a simple form by using the generalised flow equation with emergent composites as introduced in \cite{Pawlowski:2005xe}. 

For details we refer to the works mentioned above, here we briefly recapitulate the PIRG approach in \Cref{sec:PIRGs} and the ground state expansion in \Cref{sec:GroundState}.

\subsection{Physics-informed RG flows}
\label{sec:PIRGs}

PIRGs are based on generalised flow equations for the generating functional at hand, including not only RG steps such as the integration of momentum shells but also general reparametrisations of the theory, see \cite{Ihssen:2024ihp}. For the Wilsonian effective action or path integral measure the general RG setup has been constructed in \cite{Wegner_1974}, leading to the Wegner equation. Its simplest form without reparametrisations of the fields is the Polchinski equation \cite{Polchinski1984}. For the one-particle irreducible (1PI) effective action the generalised flow equation has been constructed in \cite{Pawlowski:2005xe}, its simplest form without reparametrisations is the Wetterich equation \cite{Wetterich:1992yh}. Both, the Polchinski equation and the Wetterich equation can be considered as benchmark (or baseline) equations and we include results from the latter and its proper-time variant as the baseline here. 

The qualitatively novel ingredient of the PIRG setup is the full use of reparametrisations of the theory: instead of considering the effective action $\Gamma_\varphi[\varphi]$ of the fundamental field displayed in \labelcref{eq:Gamma1stOrder}, we consider the effective action of a composite operator (mean) field 
\begin{align}
\phi=\langle \hat \phi[\hat \varphi]\rangle \,, 
\end{align}
where $\hat\phi$ is introduced to the path integral with a respective current term. In short, instead of resolving the effective action for a given fundamental mean field $\varphi$, we resolve the pair 
\begin{align}
	\left(\Gamma_\phi[\phi], \phi[\varphi]\right) \,. 
\label{eq:PIpair}
\end{align} 
This gives us the maximal freedom for performing computations. Note that $\Gamma_\phi[\phi]$ is not simply a reparametrisation of the standard effective action: it only agrees with the latter on the solution of the equations of motion of $\varphi$ and $\phi$, for more details see \cite{Ihssen:2024ihp}. For example, this novel view on functional flows allows us to completely revert the common starting point: instead of starting with the definition of the composite field operator $\phi$, we may start with choosing a specific effective action, the \textit{target action} $\Gamma_T[\phi] =\Gamma_\phi[\phi]$ and compute the respective composite operator.  

With functional flow equations one recasts the task of solving the path integral in solving it differentially, typically in terms of momentum shells with the momentum cutoff scale $k$. Accordingly, the flow of the target action and the respective composite field  
\begin{align}
	\partial_t \Gamma_\phi \equiv \partial_t \Gamma_T \,,\qquad \dot \phi = \langle \partial_t \hat \phi_k[\hat \varphi] \rangle \,,
\label{eq:targetAction}
\end{align}
is governed by the generalised flow for the 1PI effective action \cite{Pawlowski:2005xe}, 
\begin{align}
		\left[ \partial_t + \int_\tau \dot{\phi} \frac{\delta}{\delta \phi} \right]\Gamma_\phi[\phi] =\frac{1}{2} \Tr\left[ G_{\phi}[\phi]\left(\partial_t + 2 \frac{\delta \dot{\phi}}{\delta \phi} \right) R_k\right], 
		\label{eq:GenFlow} 
\end{align}
where the trace is evaluated over momentum space. The dimensionless RG time is given by $t = \ln (k/k_{\mathrm{ref}})$, where $k_{\mathrm{ref}}$ is some reference scale. Typically the initial scale $\Lambda$ is chosen. The corresponding propagator $G_{\phi}=\langle \hat\phi(\omega) \hat\phi(\nu)\rangle_c$ of the composite field $\phi$ reads 
\begin{align}
	G_{\phi}[\phi](\omega, \nu) = \left[ \frac{1}{\Gamma_\phi^{(2)}[\phi]+R_k}\right](\omega, \nu)\,,
	\label{eq:Prop}
\end{align}
with $\Gamma_\phi^{(n)}= \delta^n\Gamma_\phi/{\delta\phi^n}$. The Wetterich flow \cite{Wetterich:1992yh} is recovered from \labelcref{eq:GenFlow} with $\dot \phi= 0$, that is $\phi= \varphi$. The field transformation can be reconstructed on the level of the mean fields. To that end we use the relation
\begin{align} 
	\phi[\varphi]= \varphi - \int^\Lambda_0 \frac{d k}{k} \dot \phi[\phi_{k}]\,. 
	\label{eq:phi-varphi}
\end{align} 
\Cref{eq:phi-varphi} relates the original, fundamental (mean) field $\varphi$ to the composite field $\phi$. Note that this does not enable us to learn the transformation on the level of the field operators or their flow $\partial_t \hat\phi[\hat\varphi]$. In general this leads to the additional task of reconstructing correlation functions of the fundamental field. This was discussed and solved in \cite{Ihssen:2024ihp} at the example of a simple integral. In the present context, this route is expanded on further in \Cref{app:Reconstruction}. In any case, for $\phi[\varphi]\neq \varphi$ and hence $\hat\phi[\hat\varphi]\neq\hat\varphi$, the effective action of the composite fields is not that of the fundamental field, 
\begin{align}
	\phi \neq \varphi \implies \Gamma_\varphi[\varphi] \neq \Gamma_\phi\bigl[\phi[\varphi]\bigr]\,. 
\label{eq:GammaNotGamma}
\end{align}
This originates in the fact that the Legendre transformation is taken with respect to different field operators, see also \cite{Ihssen:2024ihp, Ihssen:2025cff} for explicit examples. 

In the present work we use a specific expansion scheme in PIRG flows, the ground state expansion discussed below in \Cref{sec:GroundState}, augmented with a derivative expansion. It is suggestive that within this combined scheme the differences \labelcref{eq:GammaNotGamma} between the effective actions are minimal and the dominant feature of this PIRG scheme is an optimisation of the convergence of the derivative expansion and its numerical optimisation. We shall test this educated guess within the computation of the energy gap in the first order derivative expansion, 
\begin{align}
	\Gamma_\varphi(\varphi)\approx \Gamma_\phi\left(\phi(\varphi)\right)\,. 
	\label{eq:Gammavarphi=Gamma_phi}
\end{align}
This additional approximation is evaluated in \Cref{app:Reconstruction} in light of the exact but more involved complete reconstruction scheme derived in \cite{Ihssen:2024ihp}. 

Even though this approximation is shown to work relatively well, we shall not use it for our analysis of the instanton-dominated regime. Instead we devise a new observable that is tailor-made for predicting the prefactor 
$a_\textrm{inst}$ in \labelcref{eq:DeltaETop}, see \Cref{sec:ExponentialPIRG}.

\subsection{Ground state expansion}
\label{sec:GroundState}
	
In the present setup, we use the \textit{target action} approach \labelcref{eq:PIpair,eq:targetAction} to reduce the dispersion relation to a classical one. This absorbs the wave function $Z_\varphi[\varphi]$ of the fundamental field $\varphi$ into the composite field $\phi[\varphi]$ with
\begin{align} 
	Z_\phi[\phi] \equiv 1\,.
\label{eq:Zphi}
\end{align}
\Cref{eq:Zphi} implements a classical dispersion relation. All fluctuation physics $\propto p^2$ is absorbed into the field definition, and hence $Z_\varphi(\varphi)$ is absorbed in $\phi$. Accordingly, the target action in the first order derivative expansion of the PIRG is given by the leading order derivative expansion,
\begin{align}
	\Gamma_{T,k}[\phi] = \int d \tau \left\{ \frac{1}{2} (\partial_\mu \phi)^2 + V_{\phi,k}(\phi) + \mathcal{O}(\partial^4) \right\}\,. 
	\label{eq:Taction}
\end{align}
This expansion has already been investigated in detail in O(N) theories in \cite{Ihssen:2023nqd} with very promising results. It is called a ground state expansion, because it allows for the interpretation of $\phi$ as the field associated to this physical state with the classical (on-shell) dispersion. 

In the present work \labelcref{eq:Zphi} is implemented with a map $\dot\phi(\phi)$ that does not contain any derivatives, see \labelcref{eq:dotphi} below. Accordingly, the first order derivative terms in $\Gamma_\varphi$ and any observables derived from it, can only originate in the classical dispersion in $\Gamma_\phi$ and the map $\phi(\varphi)$. This also entails that the present expansion scheme supports the approximate identification \labelcref{eq:Gammavarphi=Gamma_phi}. In combination this leads to approximate relation 
\begin{align} 
	\int d\tau \,\frac12 Z_\varphi(\varphi) (\partial_\mu \varphi)^2 \approx \int d\tau \,\frac12 \Bigl(\partial_\mu \phi(\varphi)\Bigr)^2 \,,
	\label{eq:Approxphivarphi}
\end{align}
which implies 
\begin{align}
	Z_\varphi(\varphi) \approx \left[ \phi'(\varphi)\right]^2\,, 
	\label{eq:Zvarphiphi}
\end{align}
up to higher order terms. \Cref{eq:Zvarphiphi} is already suggested by the dispersion \labelcref{eq:Zphi} itself.

On a technical level, the ground state expansion leads to a vanishing flow of $Z_\phi$ in \labelcref{eq:GenFlow} with the target action $\Gamma_T=\Gamma_\phi$, 
\begin{align}
	\partial_t Z_{\phi,k}(\phi) =\left. \partial_{\omega^2} \frac{\delta^2 \partial_t \Gamma_{T,k}[\phi]}{\delta \phi(-\omega) \delta \phi(\omega)} \right|_{\omega=0} \overset{!}{=} 0 \,.
\label{eq:Zphi=1}
\end{align}
The effective potential $V_{\phi,k}(\phi)$ of the composite field $\phi$ is determined by the dynamics of the regulator induced flow, i.e.~the right-hand side of \labelcref{eq:GenFlow}. The target action \labelcref{eq:Taction} implements the classical frequency dependence of the two-point function
\begin{align}
		\Gamma_{k}^{(2)}(\omega,-\omega)[\phi] = \omega^2 + \partial_\phi^2 V_{\phi, k}(\phi) \,,
\label{eq:classDisp}
\end{align}
and removes all frequency dependences of higher order vertex functions $\Gamma_k^{(n)}[\phi]$ with $n>2$ at the first order of the derivative expansion. This is a significant simplification in comparison to the Wetterich setup, see \labelcref{eq:highervertices}.

It is left to determine the flowing field leading to the target action \labelcref{eq:Taction}, which we parametrise as
\begin{align}
		\dot \phi = - \frac{f_k(\phi)}{2} \,.
\label{eq:dotphi}
\end{align}
The pair \labelcref{eq:PIpair} is computed from the flow equation for the effective potential $V_\phi(\phi)$ and that of the flowing field transformation $f_k(\phi)$. 
Inserting the above parametrisations in the generalised flow \labelcref{eq:GenFlow} for constant $\phi(\tau)=\phi$ leads us to 
\begin{subequations}
\label{eq:dtFull}
\begin{align}
			\partial_t& V_\phi = \frac{A_d k^{d+2}}{k^2 + V_\phi^{(2)}(\phi)} \left(1 - \frac{f^{(1)}(\phi)}{d+2}\right) + \frac12 f(\phi) \ V_\phi^{(1)}(\phi)\,, 
\label{eq:dtVphi} 
\end{align}
where we have dropped the $k$ dependence of $V_\phi$ and $f$ for the sake of visibility and
\begin{align}\label{eq:Ad}
	A_{d} = \frac{2 \pi^{d/2} }{ (2 \pi)^{d}\Gamma(d/2) d} \,.
\end{align}
We have used the flat regulator in the evaluation of the flow equation, see \Cref{app:reg} for more details.
The derivative of the reparametrisation function $f$ satisfies the algebraic relation 
\begin{align}
f^{(1)}(\phi)= A_d k^{d+2}\frac{ \left[V_\phi^{(3)}(\phi)\right]^2}{\left[k^2 + V_\phi^{(2)}(\phi)\right]^4} \,,
\label{eq:dtf1}
\end{align}
\end{subequations}
which follows from the constraint \labelcref{eq:Zphi=1}. The integration constant is adjusted as 
\begin{align}\label{eq:bdcond}
	f(0)= 0 \,, 
\end{align}
which implements the $\phi \to - \phi$ symmetry for the flowing field. Moreover, it also leads to \textit{local} field transformations, i.e.~
\begin{align}
	\lim_{|\phi| \to \infty} \dot \phi = \mathrm{const} \,.
\end{align}
The locality of the field transformation is pivotal for the formal existence of the emergent composite, for a detailed discussion see \cite{Ihssen:2024ihp}.

\section{The effective potential}
\label{sec:Numerics}

This Section details the evaluation of the flow of the effective potential $V_\phi(\rho)$. We discuss the numerical setup in \Cref{sec:implementation}. In particular, we introduce an approximation of third order derivative terms of the potential, that allows for an easy implementation of a numerically stable scheme. This discussion is followed by one of the shape of the effective potential in \Cref{sec:ExponentiallyFlatRegime}, specifically concentrating on the exponentially flat regime in the effective potential.

\subsection{Numerical evaluation and approximation} 
\label{sec:implementation}

We solve the flow equation of the fully field-dependent potential \labelcref{eq:dtVphi} numerically, using a continuous Galerkin method for the discretisation of field-space. We make use of the numerical framework DiFfRG \cite{Sattler:2024ozv}, which was first used in \cite{Ihssen:2023xlp}. This is detailed further in \Cref{app:Numerics}, where we also briefly discuss the fluid-dynamic approach to solving RG flows. At every step of the RG time evolution, we integrate $f^{(1)}$ as given by \labelcref{eq:dtf1} together with the boundary condition \labelcref{eq:bdcond}. This resolves the field reparametrisation \labelcref{eq:dotphi}. It is a feature of the current truncation and the scalar theory with one component, that $f^{(1)}$ is given algebraically.

The diagrammatic part of the flow of the wave function \labelcref{eq:dtf1} contains a third derivative, which is numerically hard to accommodate for in a fully field-dependent setup, built on the basis of convection (first derivatives) and diffusion (second derivatives) dynamics. This term also appears in the first order derivative expansion of the standard Wetterich flows, which are briefly discussed in \Cref{app:Wetterich1stod}, and is not exclusive to the PIRG setup.
A formulation of a stable numerical scheme including these terms is beyond the scope of the present work. 

Instead, we work with a numerically motivated approximation for this term. To begin with, we use the fluid-dynamical code frameworks \cite{Grossi:2019urj, Grossi:2021ksl, Koenigstein:2021rxj, Koenigstein:2021syz, Steil:2021cbu, Ihssen:2022xkr, Ihssen:2023qaq,Ihssen:2023xlp,Koenigstein:2023yzv, Ihssen:2024miv, Sattler:2024ozv, Zorbach:2024rre} that are formulated in the scalar invariant 
\begin{align}
	\rho = \frac{\phi^2}{2}\,. 
\end{align}
These frameworks are described in more details in \Cref{app:Numerics}. The stability of the numerics is ensured by approximating the third derivative as
\begin{align}\notag
	\left[V_\phi^{(3)}(\rho)\right]^2 &= 2 \rho\, \Bigl[3  V_\phi^{\prime\prime}(\rho)+ 2 \rho\, V_\phi^{\prime\prime\prime}(\rho)\Bigr]^2\\[1ex]
	&\approx 2 \rho \, \Bigl[3  V_\phi^{\prime\prime}(\rho)+ 2 \rho\, V_\phi^{\prime\prime\prime}(\rho=0)\Bigr]^2 \,,
\label{eq:approx1}
\end{align}
where we use the notation $V^\prime =\partial_\rho V$, and similarly for higher order derivatives w.r.t.~$\rho$. $V_\phi^{\prime\prime\prime}(\rho=0)$ refers to the value of the third derivative on the equations of motion at $\rho=0$. 

Note, that the approximation in \labelcref{eq:approx1} still accommodates a sizeable part of all higher order scatterings in the field-dependent propagator and also the vertex itself via $\partial^2_\rho V_\phi$. 
However, it does not capture the non-linear dynamics created by the $\partial_\rho^3 V_\phi$ term.

\begin{figure*}
	\centering
	\begin{subfigure}{.48\linewidth}
		\centering
		\includegraphics[width=\linewidth]{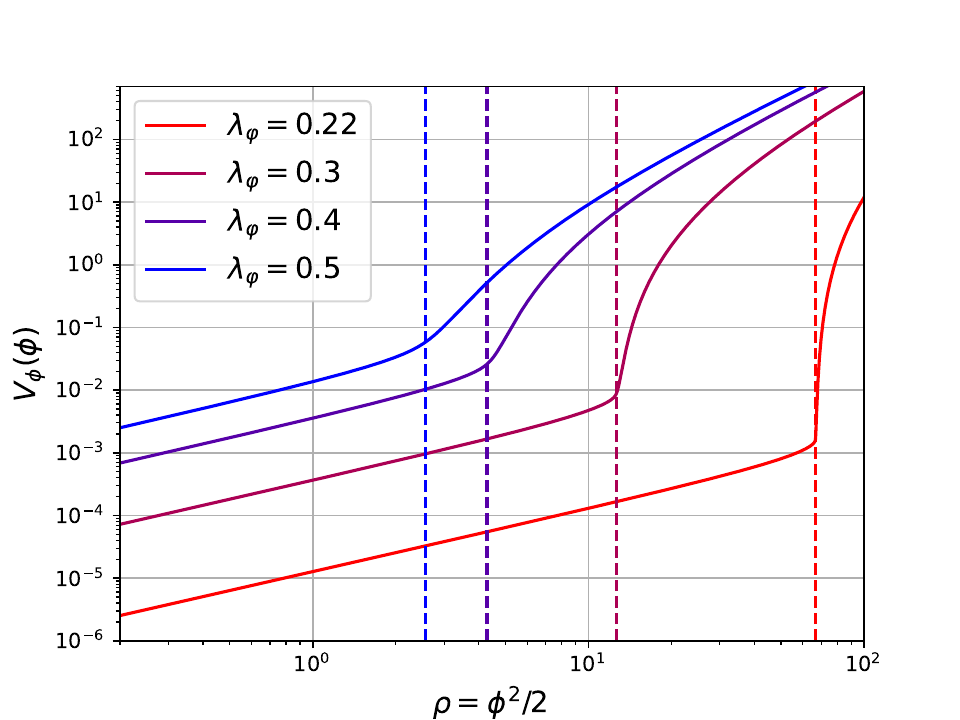}
		\subcaption{Potential $V_\phi(\phi)$.\hspace*{\fill}}
		\label{fig:Potentials} 
	\end{subfigure}%
	\hspace{0.02\linewidth}%
	\begin{subfigure}{.48\linewidth}
		\centering
		\includegraphics[width=\linewidth]{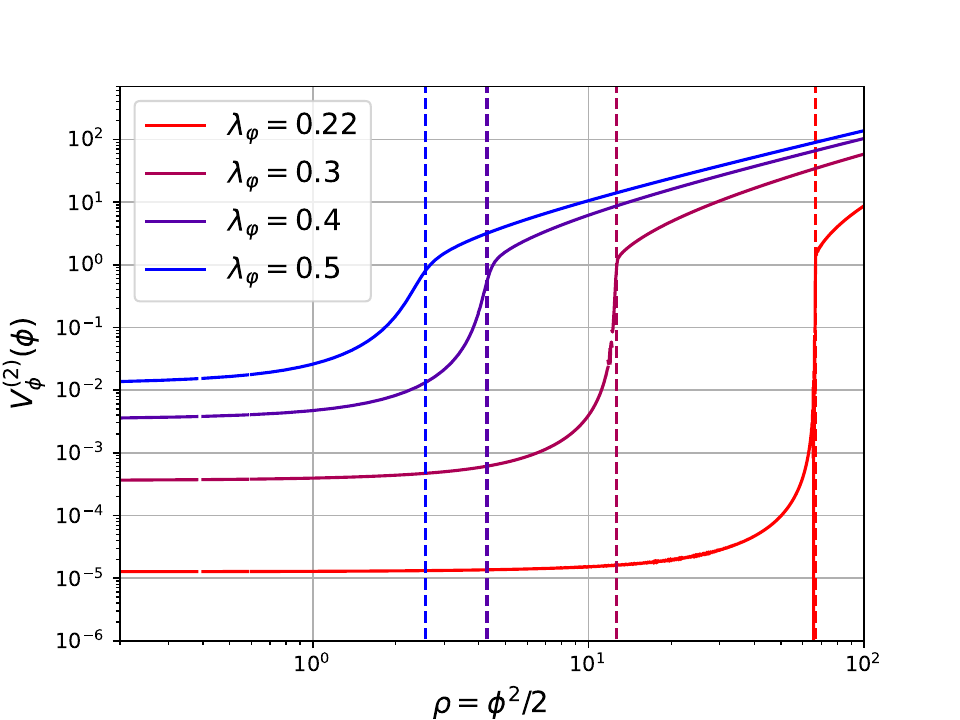}
		\subcaption{Curvature $\partial_\phi^2 V_\phi(\phi)$.\hspace*{\fill}}
		\label{fig:ddV}
	\end{subfigure}%
	\caption{Field dependence on the composite field $\phi$ for different values of the coupling $\lambda_\varphi$. The exponentially flat regime increases exponentially with $\lambda_\varphi\to 0$ as noted in \labelcref{eq:ExponentialRegime}. We have indicated the end of the exponentially flat regime, defined by $\rho^{\ }_c$ in \labelcref{eq:EndofFlat}, with dashed vertical lines.\hspace*{\fill}}
	\label{fig:Potentials2}
\end{figure*}
%
\subsection{The exponentially flat regime}
\label{sec:ExponentiallyFlatRegime}

The computation is initialised at a UV-cutoff scale $\Lambda = 300$ in units of the mass This is sufficiently large to lead to fully converged results, if using only the terms in the classical action in the initial effective action. The respective initial potential is given by 
\begin{align}
	V_{\phi,\Lambda}(\phi) = -\frac{1}{2}\phi^2 + \frac{\lambda_\varphi}{8} \phi^4 \,.
\label{eq:iniConditions}
\end{align}
Moreover, at the initial scale $\Lambda$ we identify the composite mean field $\phi$ with the fundamental mean field $\varphi$,
\begin{align} 
	\phi_\Lambda[\varphi] = \varphi\,. 
\label{eq:PhiLambda} 
\end{align}
Hence, the initial effective potential of the composite field agrees with that of the fundamental field, 
\begin{align} 
	V_{\phi,\Lambda}(\phi) = V_{\varphi,\Lambda}(\varphi)\,.
	\label{eq:iniConditionCompositeFundamental}
\end{align}
This setup leaves us with one tuning parameter $\lambda_\varphi$. We have checked that the results do not dependent on the UV-cutoff scale $\Lambda$, if the latter is increased while keeping $\lambda_\varphi$ fixed. This confirms numerically the above statement that $\Lambda=300$ is sufficiently large. Accordingly, together with \labelcref{eq:iniConditionCompositeFundamental} this allows us to identify the initial coupling $\lambda_\varphi$ with the classical one that we use in the Hamiltonian numerical approach for computing the energy gap $\Delta E$, up to some rescaling of the units
\begin{align}
	|m_\varphi| \to (1 + \epsilon) |m_\varphi| \,.
\end{align}
Presently, we use a rescaling of $\epsilon = 4 \times 10^{-3}$ for the values indicated in \Cref{fig:Instanton+Schroedinger+PIRG}.

In the following we present results for
\begin{align} 
	0.22 \leq \lambda_\varphi \leq 2.0 \,.
\label{eq:RangeInitialCoupling}
\end{align}
While the numerical evaluation at couplings lower than $\lambda_\varphi = 0.22$ is stable, the computational effort increases exponentially, see \Cref{app:NumDetails} for a more detailed discussion. Hence \labelcref{eq:RangeInitialCoupling} is a reasonable compromise between computational aspects and the necessity to explore a significant part of the instanton dominated region, which is located at $\lambda_\varphi \lesssim 0.4$, see \Cref{fig:Instanton+Schroedinger+PIRG}. 

The physical potential at $k \to 0$ is depicted in \Cref{fig:Potentials2} for different values of $\lambda_\varphi$. 
The pseudo-flattening behaviour of the curvature is discernable as a very large, flat regime at low field values in \Cref{fig:ddV}. The vertical lines signal the end of the flat regime at $\rho^{\ }_c$. We define them as the point of the curvature change of the logarithm of $V^{(2)}_\phi$, 
\begin{align}
	0=\partial_\rho^2 \log \left[ \partial_\rho V_\phi (\rho) \right]\vert_{\rho = \rho_c} \,.
\label{eq:EndofFlat}
\end{align}

This regime emerges very similarly to the flattening of the potential associated to spontaneous symmetry breaking in higher dimensions $d\geq 2$ and faces the same numerical challenges throughout the RG time evolution. These challenges are met with dedicated solvers, see e.g.~\cite{Sattler:2024ozv, Ihssen:2022xkr, Ihssen:2023qaq}. In contrast to $d\geq 2$, symmetry is restored in $d=1$ shortly before reaching $k \to 0$, necessitating efficient implicit time stepping schemes \cite{Ihssen:2023qaq} and integration to very high RG times. 
Presently, the numerical integration of the RG scale is performed up to $k = 10^{-6}$ which corresponds to an RG time of $t=20$.
 
In contrast to the standard Wetterich approach, the flattening behaviour is modified by the introduction of the field reparametrisation. This shows in the increase in size of the flat regime, whose boundary is indicated by the dashed lines in \Cref{fig:Potentials2}. In fact, the regime blows up exponentially and shows similar scaling to the instanton solution. The size of this regime is an important observable and we discuss it further in \Cref{sec:ResultsInstantons}.

Finally we remark that up until now, solving the Wetterich equation deep in the instanton regime using the first order of the derivative expansion was unsuccessful. We suspect this is linked to the exponential increase of numerical precision that is needed for an approach without field transformations.

\begin{figure*}
	\centering
	\begin{subfigure}{.48\linewidth}
		\centering
		\includegraphics[width=\textwidth]{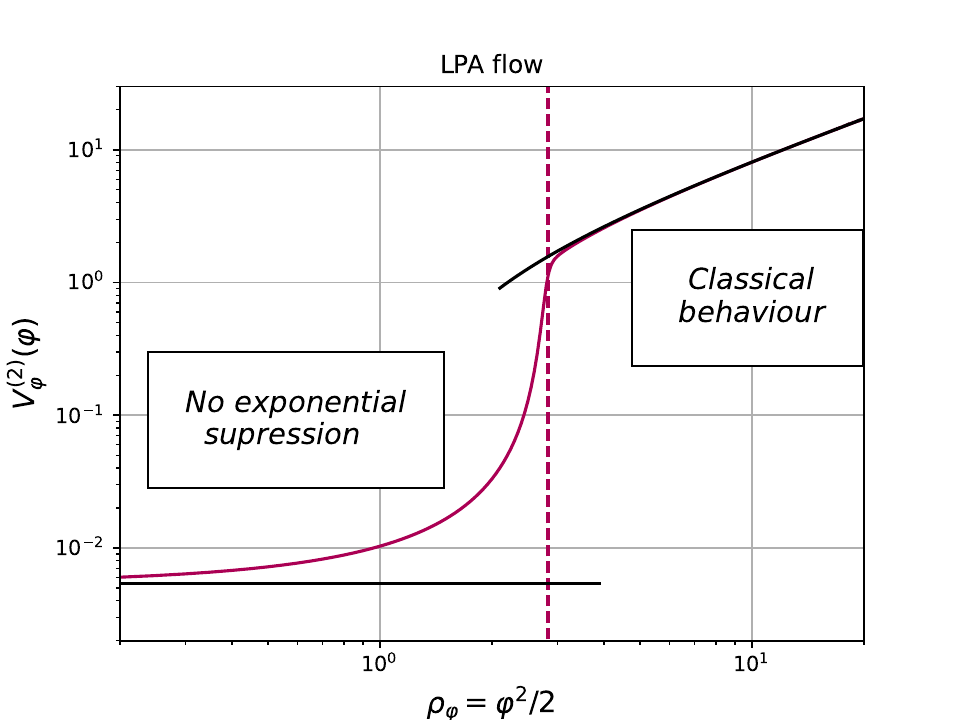}
	\end{subfigure}%
	\hspace{0.02\linewidth}%
	\begin{subfigure}{.48\linewidth}
		\centering
		\includegraphics[width=\textwidth]{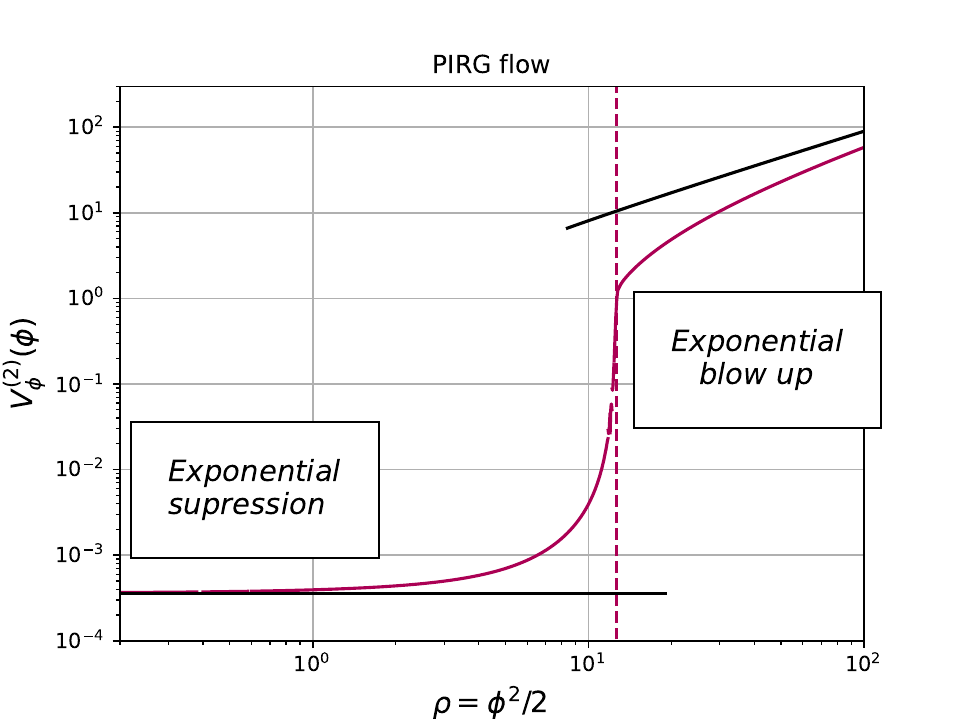}
	\end{subfigure}%
	\caption{Schematic depiction of the exponentially suppressed/enhanced quantities in both the zeroth order derivative expansion (LPA, left) and the current ground state expansion (PIRG, right) for $\lambda_\varphi=0.3$. The purple line indicates the end of the flat regime at $\rho_c$,  \labelcref{eq:EndofFlat}, for the potential $V_\phi$.\hspace*{\fill}}
	\label{fig:SchematicPlot} 
\end{figure*}
%
\section{Topology with PIRGs}
\label{sec:InstantonRegime}

In this Section we provide and discuss results in the first order derivative expansion of the PIRG ground state approach. We emphasise again that this expansion, while related to the standard derivative expansion for the Wetterich equation, is an optimised expansion scheme in terms of its convergence as well as its numerical stability, see the discussion in \Cref{sec:flowingField} and \cite{Ihssen:2024ihp}. Specifically we compute the prefactor $a_\textrm{inst}$ of the topological scaling of $\Delta E$ for $\lambda_\phi\to 0$, see \labelcref{eq:DefaInst}, which serves as a smoking gun signal for the topological tunnelling effects.

The full energy gap \labelcref{eq:DeltaE}, including the prefactors, can also be computed within the PIRG approach, but requires further preparations: to begin with, it can be computed directly from the full two-point function $\Gamma^{(2)}_\varphi(\omega)$, see \labelcref{eq:DeltaE}. In particular, it is not directly related to correlation functions of $\Gamma_\phi$, and it can only be extracted via a reconstruction, see \cite{Ihssen:2024ihp}. For the sake of completeness we discuss the respective reconstruction procedure in \Cref{app:EnergyGapPIRG}. There, we also provide results for the energy gap within a crude approximation, and discuss its viability. A full reconstruction of the energy gap in the present approach will be presented elsewhere. 

The present work pursues a different and more promising route: First we concentrate on the qualitative question whether the present fRG approach with PIRGs accommodates the instanton-induced physics effects. The presence of these topological effects is signalled by the exponential flattening of the energy gap. This exponential flattening is directly related to the exponential widening of the flat regime in the effective potential about vanishing field. In \Cref{sec:ResultsInstantons} we provide a comprehensive discussion of quantities which contain exponential scaling in the instanton-dominated regime. This allows us to dissect the numerical and structural mechanisms behind the emergence of instanton-induced effects in the fRG-approach. 

In a second step we extract the prefactor of the exponential flattening of the energy gap from the widening of the flat region, see \Cref{sec:ExponentialPIRG}.

\subsection{Dissection of the instanton-dominated regime}
\label{sec:ResultsInstantons}

We start our analysis with the observation that for $\lambda_\varphi \to 0$, the effective potential $V_\varphi(\varphi)$ of the fundamental field is necessarily exponentially flat between the minima $\pm \varphi_0$, \labelcref{eq:varphi0Classical}, of the classical potential in \labelcref{eq:ClassicalAction1-ON}. For field values $\varphi^2 \gtrsim \varphi_0^2$, the effective potential is polynomial. This entails that $\varphi_c \approx \varphi_0$ in this limit. From the technical point of view, the pulling force of the flow in the non-convex regime is to weak to move the end of the flat regime, \labelcref{eq:EndofFlat}, sizably towards smaller values: the curvature $V_\varphi^{(2)}$ of the potential at $\varphi_c$ is positive and the pulling force is proportional to $\lambda_\varphi$. Accordingly it tends towards zero for $\lambda_\varphi\to 0$. In summary, the exponential flatness of the potential is a property of the flat regime $\varphi \in \left( - \varphi_c\,,\,\varphi_c\right)$ with $\varphi_c \approx \varphi_0$ and is mainly driven by the wave function $Z_\varphi$.

\subsubsection{Exponential scaling in the ground state expansion} 
\label{sec:ExponentialGroundState}

In the ground state expansion we absorb the wave function into the field $\phi$ and use an LPA-type potential as the target action of our expansion \labelcref{eq:Taction}. The absence of instanton effects in LPA implies that within this expansion, \textit{all} instanton-induced physics are necessarily sourced by the wave function $Z_\phi$ and are transmitted to the potential via the flowing composite. In \Cref{fig:SchematicPlot} we indicate all quantities in $V_\phi$ that contain exponential scaling. We also emphasise in this context that while the ground state expansion is defined by absorbing the wave function into the field, the wave function $Z_\phi$ is not simply $Z_\varphi$ but accommodates further frequency-dependent couplings of higher order by the iteration process in the flow. 
 
In summary, this leaves us with the following scenario: the exponential flattening is sourced entirely by the map $\phi(\varphi)$, and it is this scaling that enters 
the reconstruction of $\Delta E$ from the present PIRG setup discussed in \Cref{app:Reconstruction}. This suggests the relation between the exponentially scaling quantities
\begin{align} 
	\Delta E =A_c \, \rho_c^{-d_c}+ \textrm{subleading} \,.
\label{eq:DeltaErho} 
\end{align}   
The parameters $A_c,d_c$ in \labelcref{eq:DeltaErho} are fixed in the asymptotic perturbative regime with $1.50\lesssim \lambda_\varphi \lesssim 1.86$, see \labelcref{eq:largelambda} and \Cref{fig:highfit} in \Cref{sec:ExponentialPIRG}. Values for these parameters in the present approximation are provided in \labelcref{eq:dc+ec}. Crucially, the relation \labelcref{eq:DeltaErho} carries no topological information. 

In conclusion, the location of the end of the exponential regime $\rho_c$ defined in \labelcref{eq:DefaInst} has to scale  exponentially with  
\begin{align} 
	\rho_c^{-d_c} \propto e^{-\frac{4 \sqrt{2} m_\varphi^3}{3 \lambda_\varphi}}\,. 
		 \label{eq:ExponentialRegime}
\end{align}
\Cref{eq:ExponentialRegime} entails an exponential widening of the flat regime which has already been discussed in \Cref{sec:ExponentiallyFlatRegime}. Moreover, this observable has the benefit of easy accessibility and numerical stability in the present setup: it is exponentially growing and its relative error within the given numerical setup is small. In contrast, $\Delta E$ is exponentially small and numerical errors on $V_\varphi/ Z_\varphi$ or $V_\phi$ get exponentially enhanced, which requires an exponential increase of precision for $\lambda_\varphi\to 0$. 

\begin{figure*}
	\centering
	\begin{subfigure}{.48\linewidth}
		\centering
		\includegraphics[width=\linewidth]{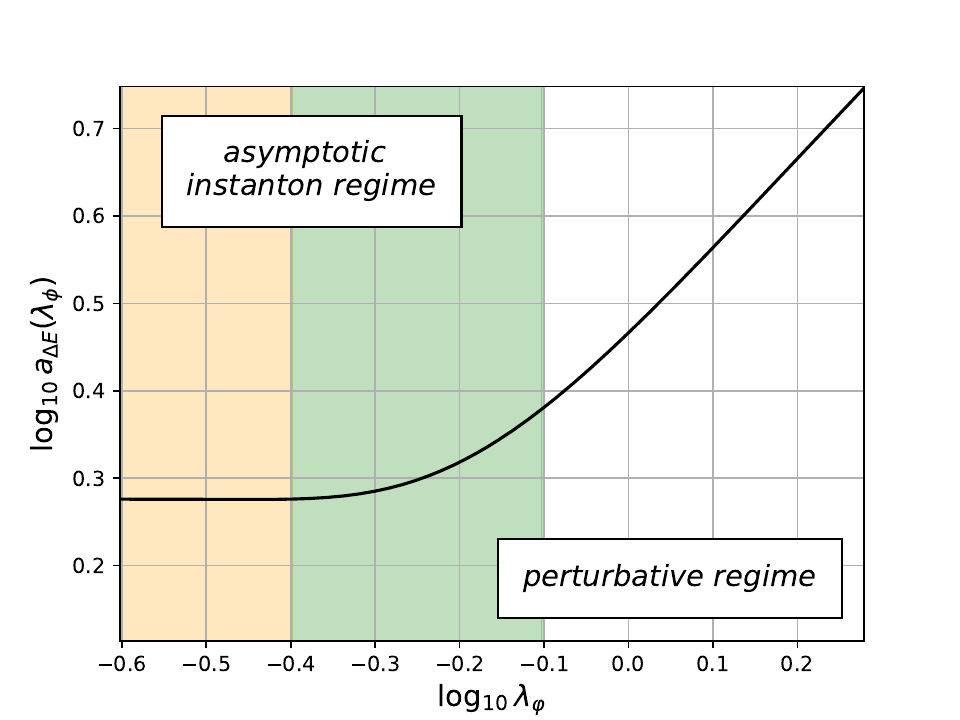}
		\subcaption{Scaling observable \labelcref{eq:a_DeltaE} of the energy gap $\Delta E$. The observable is designed such that it is flat in the instanton-dominated regime and linear in the perturbative regime. \hspace*{\fill}}
		\label{fig:Instanton}
	\end{subfigure}%
	\hspace{0.02\linewidth}%
	\begin{subfigure}{.48\linewidth}
		\centering
		\includegraphics[width=\linewidth]{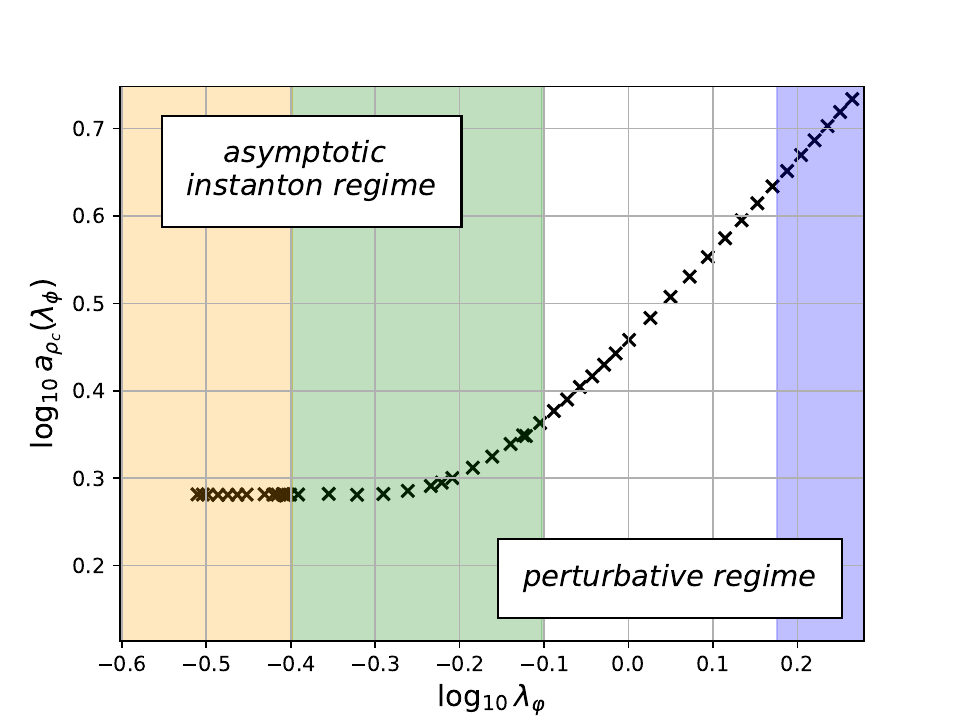}
		\subcaption{Scaling observable \labelcref{eq:a_rhoc}, of $\rho_c$, which measures the length of the flat area in the effective potential $V_\phi$, see \labelcref{eq:EndofFlat} in \Cref{sec:ExponentiallyFlatRegime}. \hspace*{\fill}}
		\label{fig:SmokingGun} 
	\end{subfigure}%
	\caption{Scaling analysis with the scaling observables $\Delta E$, \labelcref{eq:a_DeltaE}, shown in \Cref{fig:Instanton}, and $\rho_c$, \labelcref{eq:a_rhoc}, shown in \Cref{fig:SmokingGun}. We indicate three different regimes: the instanton-dominated regime (orange), the transition regime (green) and the asymptotic perturbative regime (blue). The latter is used for fixing the parameters $A_c,d_c$ in the relation \labelcref{eq:DeltaErho} between the energy gap $\Delta E$ and $\rho_c$. \hspace*{\fill}}
	\label{fig:pltos}
\end{figure*}
%

\subsubsection{Mapping out the instanton regime}
\label{sec:MappingOutInstanton}

Before proceeding with the extraction of $a_\textrm{inst}$ from the numerical data of $\rho_c$ at hand, we illustrate the procedure with our baseline solution for $\Delta E$, i.e.~the eigenvalues displayed in \Cref{fig:Instanton+Schroedinger+PIRG}. This analysis can be used to map out three different regimes for the anharmonic oscillator: 
\begin{itemize} 
\item[(1)] $\lambda_\varphi \gtrsim 0.8$:  The perturbative regime with a polynomial behaviour of all observables in $\lambda_\varphi$, and in particular of $\Delta E$ and $\rho_c$. 
\item[(2)] $0.8\gtrsim \lambda_\varphi \gtrsim 0.4$: The transition regime between the perturbative one and the instanton-driven regime. We expect that this regime is subject to intricate dynamics as it has to accommodate both, perturbative and topological effects. 
\item[(3)] $\lambda_\varphi \lesssim 0.4$: The instanton-dominated regime which is well described by a saddle point expansion about the tunnelling solution.  
\end{itemize} 
Now we construct an observable which shows a trivial scaling in the two asymptotic regimes. In particular, the polynomial sub-scaling of the instanton-dominated regime induced by the perturbative saddle point expansion about the tunnelling solution is taken into account. The observable is given by 
\begin{align}
	a_{\Delta E}(\lambda_\varphi) =  \lambda_\varphi^2 \,\partial_{\lambda_\varphi} \log \left[ \left(1 +	c_{\Delta E}^{ \ } \lambda_\varphi^2\right)\sqrt{\lambda_\varphi}\, \Delta E \right]  \,. 
\label{eq:a_DeltaE}
\end{align}
\Cref{eq:a_DeltaE} tends towards the instanton coefficient $a_\textrm{inst}$ in \labelcref{eq:DeltaETop} for $\lambda_\varphi \to 0$, 
\begin{align} 
		a_{\Delta E}(0) = a_\textrm{inst} \approx 1.886 \,.
		\label{eq:ainstSchrodinger}
\end{align}
The factor $(1+ c_{\Delta E}^{ \ }\lambda_\varphi^2)$ accommodates the sub-leading scaling in the instanton-dominated regime. It is worth noting that the linear two-loop part in \labelcref{eq:DeltaETop} leads to a quadratic term proportional to $\lambda_\varphi^2$ in \labelcref{eq:a_DeltaE}, but this term is apparently cancelled by further ones from the polynomial prefactor in $\Delta E$. We are left with the cubic subleading term from $\lambda_\varphi ^2 \partial_{\lambda_\varphi} \log ( \sqrt{\lambda_\varphi}  \Delta E)$. We fix $	c_{\Delta E}^{ \ }$ such that this contribution is cancelled, leading to 
\begin{align} 
	c_{\Delta E}^{ \ }= 1.781\,.
\end{align}
This construction leaves us with the observable $a_{\Delta E}(\lambda_\varphi)$ which is flat in the instanton-dominated regime and shows a linear running with $\lambda_\varphi$ in the perturbative regime. Clearly, the transition regime between the flat instanton-dominated and linear perturbative regime has a non-trivial $\lambda_\varphi$-dependence. The result for $a_{\Delta E}(\lambda_\varphi)$ is shown in \Cref{fig:Instanton}. We find, that the flat instanton-dominated regime (in orange) reaches to $\lambda_\varphi \approx 0.4$. The interface regime (2) (in green) is located at $0.8\gtrsim \lambda_\varphi \gtrsim 0.4$, and the perturbative regime (in white) is defined by $\lambda_\varphi\gtrsim 0.8$ with a clear linear (i.e.~purely polynomial) scaling. 

\begin{figure*}
	\centering
	\begin{subfigure}{.48\linewidth}
		\centering
		\includegraphics[width=\linewidth]{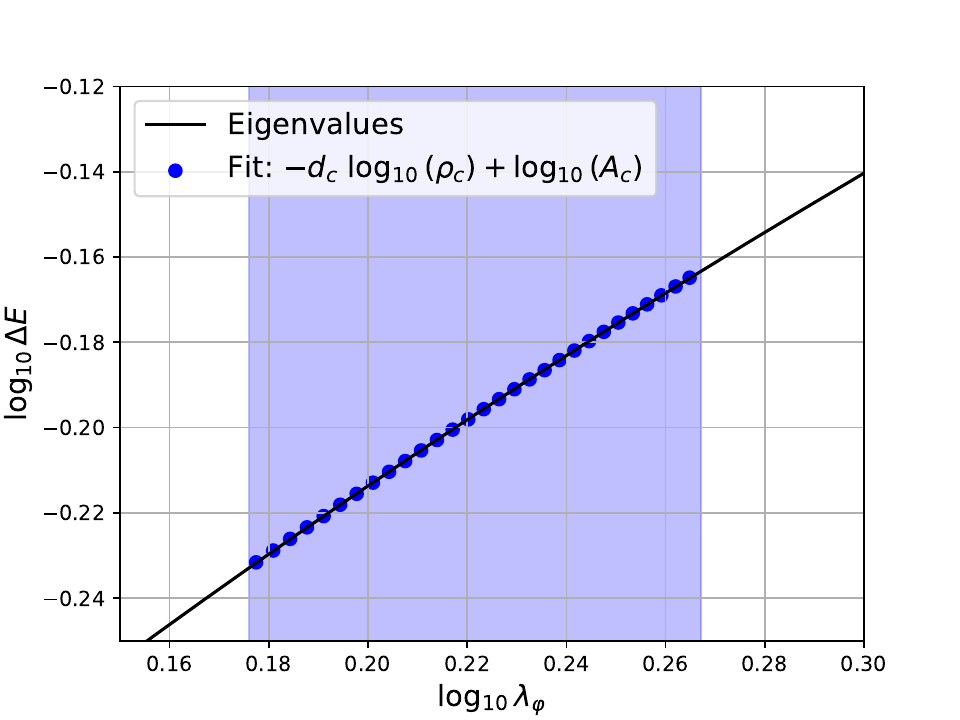}
		\subcaption{Rescaling of the new observable $\rho_c$ to match the data of the energy gap in the perturbative regime. The fit parameters are given by $d_c= 1.269(1)$ and $A_c=-0.3834(3)$.\hspace*{\fill}}
		\label{fig:highfit}
	\end{subfigure}%
	\hspace{0.02\linewidth}%
	\begin{subfigure}{.48\linewidth}
		\centering
		\includegraphics[width=\linewidth]{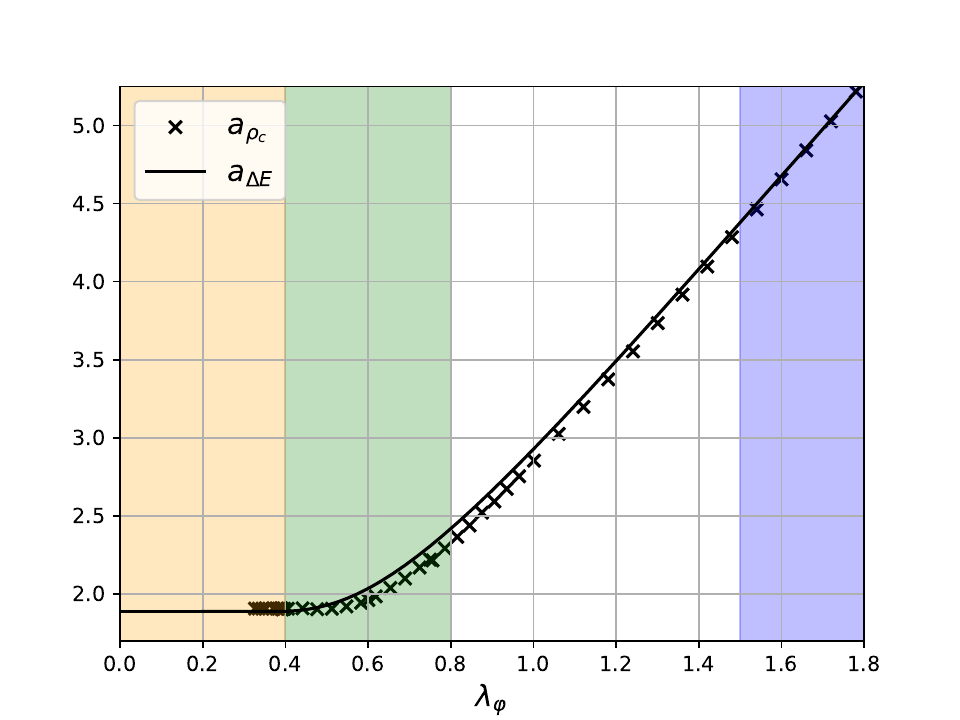}
		\subcaption{Projection of the pseudo-flat regime using \labelcref{eq:a_rhoc}, the energy gap is projected with \labelcref{eq:a_DeltaE}. We use the same shading as in \Cref{fig:pltos}.\hspace*{\fill}}
		\label{fig:arho0aDeltaEComp}   
	\end{subfigure}%
	\caption{Extrapolation of the instanton behaviour from $\rho_c$ (the size of the pseudo-flat regime). First $\rho_c$ is fitted to match $\Delta E$ in the perturbative regime (1), then we use the projection \labelcref{eq:a_rhoc} to extract the exponential suppression in the instanton regime (3). \hspace*{\fill}}
	\label{fig:arho0aDeltaE} 
\end{figure*}
%

\subsection{Topological scaling from PIRGs} 
\label{sec:ExponentialPIRG} 

The analysis in \Cref{sec:MappingOutInstanton} is readily done also with the observable $\rho^{-d_c}(\lambda_\varphi)$ instead of $\Delta E(\lambda_\varphi)$. We define 
\begin{align}
	a_{\rho_c}(\lambda_\phi) = \lambda_\varphi^2 \, \partial_{\lambda_\varphi} \, \log \left[ \left(1 + 
	c_{\rho_c}^{ \ } \lambda_\varphi^2\right)\sqrt{\lambda_\varphi} \,\rho_c^{-d_c}  \right] \,. 
	\label{eq:a_rhoc}
\end{align}
The prefactor $d_c$ takes into account the non-trivial power in the map from $\rho_c$ to $\Delta E$. It is a universal power and $d_c$ is fixed in the asymptotic perturbative regime. This is done by fitting 
\begin{align} 
	\log \Delta E(\lambda_\varphi) \approx  - d_c \log \rho_c(\lambda_\varphi) + \log A_c\,, 
\label{eq:FitPerturbative}
\end{align}
deep in the perturbative regime with 
\begin{align} 
\qquad 1.50\lesssim \lambda_\varphi \lesssim 1.86\,, 
\label{eq:largelambda}
\end{align}
see \Cref{fig:highfit}. The additional term $A_c$ takes into account the prefactor in the relation \labelcref{eq:DeltaErho}. The best $\chi^2$-fit is obtained with 
\begin{align} 
d_c= 1.269(1)\,,\qquad \log A_c=-0.3834(3)\,.
\label{eq:dc+ec} 
\end{align} 
Finally, the parameter $c^{\ }_{\rho_c}$ is adjusted such that the subleading term of $\rho_c$ in the instanton-dominated regime is cancelled and the flat regime with $a_{\rho_c}(\lambda_\varphi) \approx a_{\rho_c}(0)$ is approached earlier. Note that $c^{\ }_{\rho_c}\neq c^{\ }_{\Delta E}$ already without approximations. This leads us to 
\begin{align} 
	c^{\ }_{\rho_c} =1.79(1) \,.
	\label{eq:crho}
\end{align} 
To obtain an error estimate, we have varied the size of the fit regime by taking away points from the lower end: For $\lambda_\varphi \lesssim 0.3$ our approximation breaks down, most likely due to the incomplete treatment of $V^{(3)}$. This failure is apparent due to a sudden and sharp rise in the projection of the data using \labelcref{eq:a_rhoc}. Fortunately, this failure occurs for $\lambda_\varphi$ deep in the flat regime where the instanton asymptotics has already fully set in. We have dropped data points with $\lambda_\varphi \lesssim 0.3$ in \Cref{fig:SmokingGun}. Importantly, the respective systematic error in the determination of $c_{\rho_c}$ is very small, see \labelcref{eq:crho}. This leads to a small error in the determination of the instanton coefficient \labelcref{eq:ainstPIRG}. 

The result for $a_{\rho_c}(\lambda_\phi)$ is shown in \Cref{fig:SmokingGun} and constitutes the main computational result in the present work: it clearly shows an extended flat regime, which signals the exponential scaling of our observable. 
This scaling is the smoking gun for the instanton-dominated regime: the present computation is a first and impressive numerical confirmation of the capacity of the fRG approach to accommodate topological effects in relatively simple approximations. 
This already exciting result is combined with the numerical one for the instanton coefficient $a_\textrm{inst}$ with  
\begin{align}
	a_{\mathrm{inst},\rho_\phi} = 1.910(2)\,.
	\label{eq:ainstPIRG}
\end{align}
\Cref{eq:ainstPIRG} a deviation on the $1\%$ level from the exact result \labelcref{eq:ainstSchrodinger}. 
Here we have extrapolated the horizontal line in the flat regime to $\lambda_\varphi=0$, and the small systematic error originates from the $\lesssim 10^{-5}$ relative deviation from flatness in this regime (measured in $a_\textrm{inst}$).  

We close this Section with a few remarks. We have derived two  exciting results:\\[-2ex] 

First of all, the present analysis has proven that the fRG approach, at least in its PIRG representation, can capture topologically driven effects already in a relatively simple approximation, the first order derivative expansion. The smoking gun plot is given by \Cref{fig:arho0aDeltaEComp}: the scaling observable $a_{\rho_c}$, \labelcref{eq:a_rhoc}, is flat in the exponentially suppressed regime and the latter is in one-to-one correspondence to the presence of instanton-induced effects. The existence of the flat regime is proven in \Cref{fig:pltos}. 

Secondly, we have used the scaling observable for the computation of the scaling coefficient $a_\textrm{inst}$ in the energy gap. The result \labelcref{eq:ainstPIRG} agrees quantitatively within $1\%$ with the analytic one, see \labelcref{eq:DeltaETop}. This quantitative agreement is far better that our own, optimistic, expectations. While it certainly is related to the optimised expansion scheme in the PIRG approach, the ground state expansion, this fact still awaits a full explanation, in particular an assessment of the dynamics at work. 

We also remark in this context that the leading order map \labelcref{eq:DeltaErho}, fixed in the asymptotic perturbative regime, is working well over the whole $\lambda_\varphi$-regime shown in \Cref{fig:arho0aDeltaEComp}: both asymptotic regimes agree very well, while the transition regime with its intricate dynamics would require higher order terms in \labelcref{eq:DeltaErho}. Still, as in the case of the instanton coefficient this awaits a full explanation.  
We leave such an evaluation and improvements upon the present approximation to future work.

\section{Conclusion and outlook}\label{sec:outlook}

In the present work we have tackled the question whether topological effects can be accommodated in functional renormalisation group flows within standard expansion schemes such as the derivative expansion. This has been done within an application to the anharmonic oscillator with its relatively simple dynamics apart from the instanton-induced effects. Using the new physics-informed functional Renormalisation Group (PIRG) \cite{Ihssen:2024ihp}, we have found solid evidence for the incorporation of topological tunnelling effects. This was the case already in the relatively simple approximation used: the first order of the derivative expansion. The progress is rooted in a combination of three novel ingredients already emphasised in \Cref{sec:Introduction}:  \textit{(i)} the use of PIRG flows within an expansion about the ground state of the theory \cite{Ihssen:2023nqd}; \textit{(ii)} the use of powerful numerical techniques, see e.g.~\cite{Ihssen:2022xkr, Ihssen:2023qaq, Sattler:2024ozv}; 
\textit{(iii)} the use of specific scaling observables optimised for the detection of the exponential instanton-induced scaling. 

Results have been discussed in detail in \Cref{sec:ExponentialPIRG}. In short, the accommodation of topological effects is in one-to-one correspondence to an emergent flat regime for the new scaling observable and is clearly visible, see \Cref{fig:SmokingGun,fig:arho0aDeltaEComp}. The resulting instanton coefficient \labelcref{eq:ainstPIRG} is in within 1\% agreement with the analytic one in \labelcref{eq:ainstSchrodinger}. This quantitative agreement within 1\% is unexpected and warrants further analysis. It is certainly related to the optimised expansion scheme, but also to the relatively simple dynamics of the system apart from the topological tunnelling effects. However, there may be even more structure to it. This and further improvements will be discussed elsewhere. 

With the present approach we envisage applications to further systems and phenomena involving topological effects, a very interesting application is e.g.~that to the Berezinski–Kosterlitz–Thouless transition \cite{berezinskii1971destruction, berezinskii1972destruction, kosterlitz1973ordering, kosterlitz1974critical}. For previous work using fixed-point analyses in the fRG see~\cite{Grater:1994qx, VonGersdorff:2000kp, Jakubczyk:2014isa, Jakubczyk:2016rvr, Defenu:2017, Giachetti:2021woq}. We hope to report on these applications in the near future.

\begin{acknowledgments}
		
We thank H.~Gies, J.~Hübner, F.~Sattler, M.~Scherer and T.~Wiethe for discussions. 
This work is funded by the Deutsche Forschungsgemeinschaft (DFG, German Research Foundation) under Germany’s Excellence Strategy EXC 2181/1 - 390900948 (the Heidelberg STRUCTURES Excellence Cluster) and the Collaborative Research Centre SFB 1225 - 273811115 (ISOQUANT). 

\end{acknowledgments}
	
	\appendix
	
	\begingroup
	\allowdisplaybreaks

\section{Numerical evaluation}
\label{app:Numerics}

In this Appendix we outline the numerical evaluation of the flow equations. The flow of the effective potential $V_k(\phi)$, \labelcref{eq:dtFull}, is a partial differential equation of convection-diffusion type and is solved numerically using the (continuous) Galerkin method. This method has also been used in the context of the fRG in 
\cite{Ihssen:2023xlp, Ihssen:2024miv, Sattler:2024ozv}. Continuous Galerkin methods (CGM) are finite element methods, which are related to Discontinuous Galerkin methods (DGM). The latter have been used in the fRG in \cite{Grossi:2019urj, Grossi:2021ksl, Ihssen:2022xkr}. Additionally, finite volume implementations of the RG flows have been investigated in \cite{Koenigstein:2021rxj, Koenigstein:2021syz, Steil:2021cbu, Koenigstein:2023yzv, Zorbach:2024rre}. While DGMs are a combination of finite element and finite volume methods which allow the quantitative evaluation of shock waves, it is often reasonable to use a simpler method if the presence of shocks can be excluded. This is the case for the present investigation of the quantum anharmonic oscillator, since to our knowledge shock development has only been found in presence of spontaneous symmetry breaking and first order phase transitions in an fRG context \cite{Grossi:2021ksl}. 

The flow equation is reformulated in terms of the invariant $\rho= \frac{\phi^2}{2}$, which manifests the $\phi \to -\phi$ symmetry of O(1) theory. The derivatives of the potential are given in terms of the new function $u$ as 
\begin{align} \label{eq:potDerivs}
	u(\rho) := \partial_\rho V_\phi(\phi) &= \frac{V_\phi ^{(1)}}{\phi} \notag \,, \\[1ex]
	u(\rho) + 2 \rho \partial_\rho u(\rho) &= V_\phi^{(2)} \notag \,, \\[1ex]
	3 \partial_\rho u(\rho) + 2 \rho\partial^2_\rho u(\rho) &= \frac{V_\phi^{(3)}}{\phi} \,.
\end{align}
Furthermore, we take a $\rho$ derivative of the flow equation \labelcref{eq:dtFull}, such that it turns into a closed expression in $u$ and its higher derivatives, 
\begin{align}\label{eq:dtu}
	\partial_t u =& \partial_\rho \left[\frac12 f(\phi) \phi \, u
	+ \frac{A_d \,k^{d+2}}{ \left(k^2 + 	u + 2 \rho \partial_\rho u\right)} \left(1 - \frac{f'(\phi)}{3}\right)\right]
	\,,
\end{align}
with $A_d$ given in \labelcref{eq:Ad}.
The flowing field transformation in terms of $\rho$ is given by
\begin{align}\label{eq:fieldTraforho}
	f'(\phi) = 2 \rho A_d \, k^{d+2}  \frac{\left(3 \partial_\rho u(\rho) + 2 \rho\partial^2_\rho u(\rho)\right)^2}{\left(k^2 + 	u + 2 \rho \partial_\rho u\right)^4} \,.
\end{align}
We have commented on an approximation of $2 \rho \partial^2 _\rho u$ in the main text in \labelcref{eq:approx1}. For the computation of the energy gap in \Cref{app:EnergyGapPIRG} we also provide an error using a second approximation
\begin{align}
	\left[V_\phi^{(3)}(\rho)\right]^2  \approx& 2 \rho\, \Bigl[ 3 V_\phi^{\prime\prime}(\rho)\Bigr]^2\,.
	\label{eq:approx2}
\end{align}

\subsection{Numerics of the blow up}\label{app:NumDetails}

In \Cref{sec:Numerics}, we discussed the blow up of the exponentially flat region and the connected increase of numerical cost.

In general, the use of fluid-dynamical methods requires two properties of the grid:
\begin{itemize}
	\item It needs to be fine enough to capture the convexity restoring dynamics in the flat part.
	\item Its largest value needs to be big enough to assume a vanishing flux at the boundaries, see also \cite{Grossi:2019urj}.
\end{itemize}
In presence of the exponential blow up of the flat region, this entails that we need to use a exponentially larger grid with a fine resolution as we go to smaller $\lambda_\varphi$.
Thus the lower bound in \labelcref{eq:RangeInitialCoupling} is a (current) numerical bound, and our numerical simulations converge for values of the coupling $\lambda_\varphi \geq 0.22$. We emphasise that we have not designed a dedicated code for this purpose and this bound is readily lowered with modifications of the present code, such as coordinate transformations of the numerical grid.

To conclude, the limiting factor to access the fully field-dependent potential for couplings $\lambda_\varphi < 0.22$ is the size of the numerical grid, which is currently implemented as a finely resolved interval $\rho \in \mathcal{I}_1 = [0,70]$ and a coarsely resolved interval $\rho \in \mathcal{I}_2 = [70,100]$. Once the pseudo-flat regime enters $\mathcal{I}_2$ the numerical implementation fails.
The second interval is added to ensure the correct implementation of boundary conditions at large field values at a low numerical cost.

\section{The energy gap from PIRGs}
\label{app:EnergyGapPIRG}

Within the effective action approach, the energy gap is defined as the RG-independent curvature of the effective potential of the fundamental field at the minimum, \labelcref{eq:DeltaE}. We have argued in \Cref{sec:ExponentiallyFlatRegime,sec:GroundState}, that in the ground state expansion the effective action of the fundamental field is approximately given by that of the composites, \labelcref{eq:Gammavarphi=Gamma_phi}, if evaluated close to the minima of the effective actions. In combination this leads us to 
\begin{align}
	\Gamma_\varphi[\varphi] \approx \int d\tau \,\left[ \frac12 \Bigl( \phi'(\varphi) \partial_\mu \varphi\Bigr)^2 +V_\phi\left(\phi(\varphi)\right)\right]\,,
	\label{eq:GammavarphiApprox}
\end{align}
at vanishing cutoff scale, $k\to 0$ with the wave function $Z_\varphi(\varphi) \approx \phi'(\varphi)^2$, see \labelcref{eq:Zvarphiphi}. Consequently, the energy gap \labelcref{eq:DeltaE} satisfies the approximate relation 
\begin{align}
	\Delta E \approx \sqrt{\frac{V_\phi^{\prime\prime}(0) }{\phi^\prime (0)^2}}= \sqrt{V_\phi^{\prime\prime}(0) } \,, 
	\label{eq:DeltaEPIRG}
\end{align}
where we have used $Z_\varphi(0)=1$ with 
\begin{align}
	\phi^\prime(\varphi=0) =1\,. 
	\label{eq:phiprime1}
\end{align}
The latter is obtained from the field transformation \labelcref{eq:phi-varphi} and is depicted in \Cref{fig:Wavefunction} for different values of the coupling $\lambda_\varphi$. \Cref{fig:Wavefunction} shows the wave function $Z_\varphi(\varphi)$ for the approximation \labelcref{eq:GammavarphiApprox}.

\subsection{Direct results} 
\label{app:DirectResults}

In the present work we use \labelcref{eq:DeltaEPIRG} to determine the energy gap.
The results are depicted in \Cref{fig:lambdaDep}.
We also include results from a zeroth order derivative expansion (LPA), 
as well as data for a first order derivative expansion using a proper-time flow \cite{Bonanno:2019ukb}. The data is taken from \cite{Bonanno:2022edf}, Table II, column $\Delta E_{pt10}$. We include this data because this specific combination of flow equation and approximation produces, to this date and to our knowledge, the most quantitatively precise results obtained from the fRG. 

For a better comparison of the results, we also depict the relative error of our result in \Cref{fig:lambdaDep}. This relative error is defined in reference to the solution computed from the eigenvalues ($\Delta E_{\mathrm{Ev}}$)
\begin{align}
	\delta_{\#, \mathrm{rel}} = \frac{\Delta E_{\#}- \Delta E_{\mathrm{Ev}}}{\Delta E_{\mathrm{Ev}}} \,. 
	\label{eq:RelError}
\end{align}
\begin{figure}[t]
	\includegraphics[width=\linewidth]{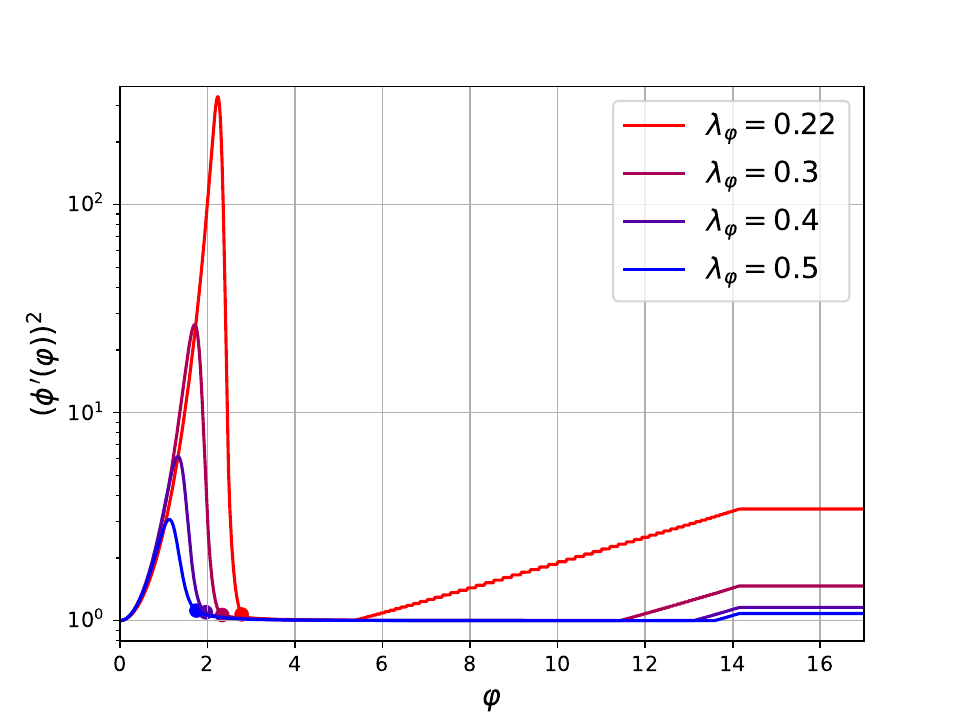}
	\caption{Field dependence of the reconstructed wave function $ Z_\varphi(\varphi)\approx (\phi'(\varphi))^2$. The end of the exponentially flat regime $ \rho_c$ is indicated by the dots. \hspace*{\fill}}
	\label{fig:Wavefunction}
\end{figure}
\begin{figure*}
	\centering
	\begin{subfigure}{.48\linewidth}
		\centering
		\includegraphics[width=\textwidth]{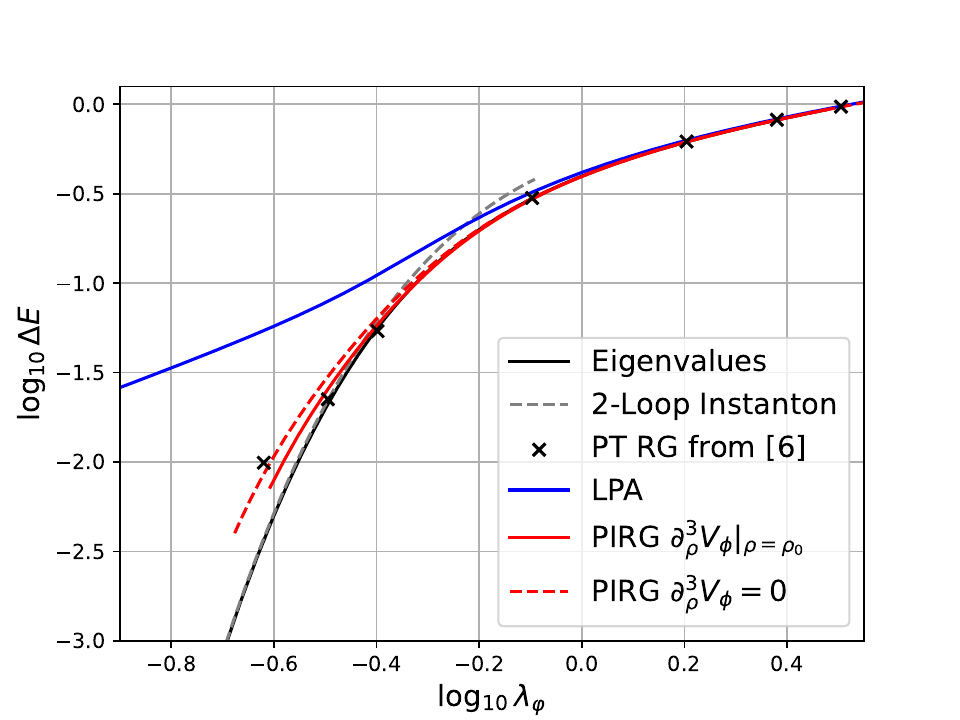}
	\end{subfigure}%
	\hspace{0.02\linewidth}%
	\begin{subfigure}{.48\linewidth}
		\centering
		\includegraphics[width=\textwidth]{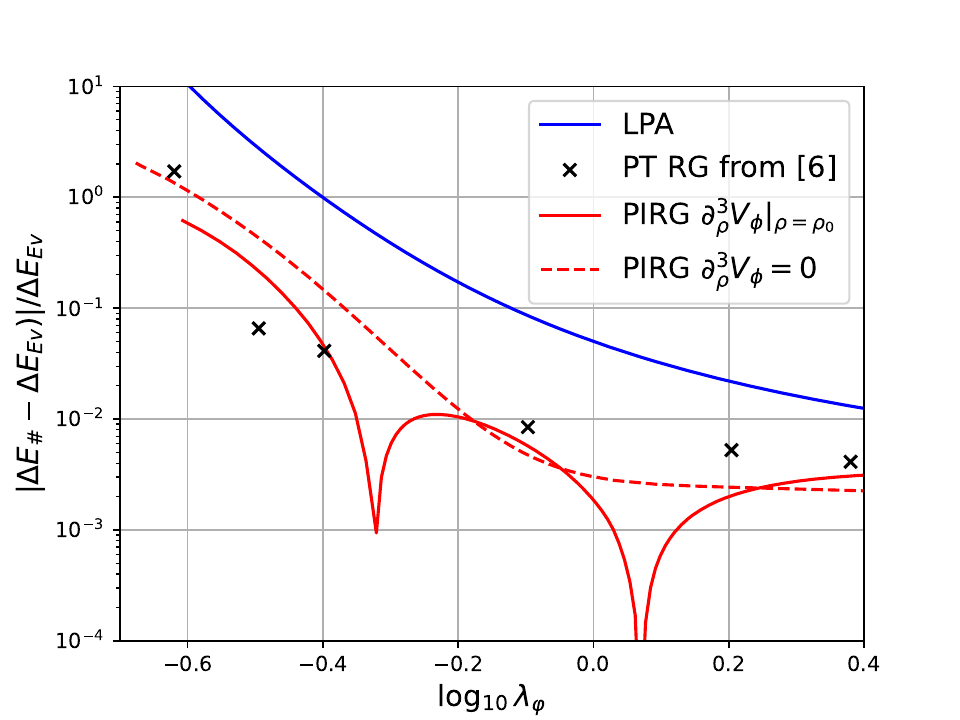}
	\end{subfigure}%
	\caption{Energy gap $\Delta E$ as a function of $\lambda_\varphi$. The eigenvalues \textit{(black)} are numerically evaluated from the Schroedinger equation and serve as a benchmark. The right plot shows the relative error \labelcref{eq:RelError} with respect to this benchmark. The zeroth order derivative expansion (LPA) calculation \textit{(blue)} is performed with $f(\phi) = 0$ and the flowing fields \textit{(red)} are evaluated using $f(\phi)$ as given by \labelcref{eq:dtf1}. The error estimate on the approximation \labelcref{eq:approx2} is dashed. Additionally we include the two-loop instanton solution \labelcref{eq:TwoLoopInst} \textit{(grey)} and reference data from \cite{Bonanno:2022edf}.\hspace*{\fill}}
	\label{fig:lambdaDep} 
\end{figure*}
%

\subsection{PIRGs and reconstruction}
\label{app:Reconstruction}

We have discussed in \Cref{sec:PIRGs,sec:GroundState}, that the identification \labelcref{eq:Gammavarphi=Gamma_phi} is only an approximation, albeit one that is supported by the ground state expansion. Still, in regimes with a strong field-dependence of the transformation special care is necessary. We know from the investigations in \cite{Ihssen:2024ihp} within the simple benchmark case of a one-dimensional integral (0+0-dimensional QFT in comparison the (1+0)-dimensional QFT for quantum mechanics), that extreme choices for the target action, such as the classical action, lead to qualitative differences in the two exact effective actions without any approximation. 

For this reason we discuss the reconstruction of observables and the effective action, which was introduced in \cite{Ihssen:2024ihp}. There it has been shown that cumulants of $\int \varphi^2$ and $\int \varphi^4$ can be computed from derivatives of the effective action $\Gamma_\phi^{(c)}[\phi]$ simply by taking derivatives w.r.t.~$m_\varphi$ and $\lambda_\varphi$. Here, the superscript ${}^{(c)}$ used in the effective action indicates a normalisation of the path integral and hence also its Legendre transform, that does not depend on the parameters of the classical action, 
\begin{align}
	\mathcal{Z}^{(c)}_\phi[J_\phi] = \int d\hat\varphi \,e^{-S[\hat\varphi] + \int d\tau\, J_\phi(\tau) \,\hat\phi[\hat\varphi]}\,, 
\end{align}
for more details see \cite{Ihssen:2024ihp}. The computation of the energy gap is more involved, since the approach does not offer access to \textit{local} correlation functions, but only to momentum (frequency) averaged quantities i.e.~
\begin{align}
	\frac{\delta \Gamma^{(c)}_\phi[0]}{\delta m^2_\varphi} = \frac12	\int_x \langle \varphi (x) \varphi(x) \rangle  =\frac12 \int_\omega \frac{\mathcal{V}_1}{ Z_\varphi(\omega) \left( \omega^2 + \Delta E^2 \right)} \,.
	\label{eq:Cumulants}
\end{align}
We have assumed the general shape $G_{\varphi \varphi}^{-1} = Z_\varphi(\omega) \left(\omega^2 + \Delta E^2\right)$ for the scalar propagator, $\Gamma[0]$ indicates the solution on the equation of motion for $\phi$, related to that of $\varphi$ with $\varphi=0$, and $\mathcal{V}_1$ is the temporal volume. \Cref{eq:Cumulants} follows from $\Gamma^{(c)}_\phi[0] =-\ln \mathcal{Z}^{(c)}_\phi[0]$.  

To proceed with the computation, we need to make an assumption for $Z_\varphi(\omega)$ in \labelcref{eq:Cumulants}: e.g.~for a constant wave function $Z_\varphi(\omega)=Z_\varphi(0)$, we obtain 
\begin{align}\label{eq:C1}
	{\cal C}_0 =\frac{\delta \Gamma^{(c)}_\phi[0]}{\delta m^2_\varphi} = \frac{\mathcal{V}_1}{4 \, Z^{1/2}_\varphi(0)\, \Delta E} \,.
\end{align}
This assumption holds true in a free theory or for positive $m^2_\varphi>0$. Using $Z_\varphi(\omega)=1$, the former can be verified analytically, the latter was confirmed numerically.

\Cref{eq:C1} already has an interesting consequence: in the PIRG framework the computation of $\Gamma_\phi$ implies a respective one of $\Gamma_\varphi$ with a given normalisation of the field. If this induced normalisation is given by $Z_\varphi(0)=1$, the integrated cumulant \labelcref{eq:C1} already provides us with the energy gap. Indeed, while being short of a proof, it is suggestive that the ground state expansion is arranging for precisely this normalisation.  

We proceed with the analysis of the general situation, where we need more information on the momentum structure of the cumulants. This information can be obtained by inserting a prefactor $Z = \left(1 \pm \epsilon\right)$ for the kinetic term in the classical action \labelcref{eq:ClassicalAction1-ON} and taking derivatives with respect to this factor instead of the couplings $m^2_\varphi, \lambda_\varphi$.

We are then able to compute cumulants of the shape
\begin{align}
	{\cal C}_n = \frac{\delta^n \Gamma^{(c)}_\phi[0]}{\delta^n Z} = - \left[-\frac12	\int_x \langle \left( \partial_\mu \varphi(x) \right)^2 \rangle \right]^n \,.
	\label{eq:Cumulantspn}
\end{align}
These integrals contain divergences of $n-1$-th order and need to be regularised with the same UV cutoff as is used in the RG flow. For example, for $n=1$ we find
\begin{align} \nonumber 
	{\cal C}_1=&	\frac{\delta \Gamma^{(c)}_\phi[0]}{\delta Z}\\[1ex]
	&\hspace{-5mm}= \mathcal{V}_1\left\{ \int_\omega \frac{\omega^2}{  Z_\varphi(\omega) \left(\omega^2 + \Delta E^2 \right)}
	- \frac{\omega^2}{  \omega^2 +R_\Lambda(\omega^2) +  m_\varphi^2}\right\}
	\,,
	\label{eq:C1prime}
\end{align}
where we have used that $\Gamma^{(c)}_\phi[0]$ is the integrated flow, beginning from the classical action at some initial scale $\Lambda$ with the corresponding regulator $R_\Lambda$.

The determination of ${\cal C}_n$ allows to make an ansatz for the the momentum dependence of $Z_\varphi(p)$ with $n$ coefficients, e.g. 
\begin{align}
	Z_\varphi(\omega) = 1 + (\Lambda - \omega)F( \omega ; c_1, \dots , c_n)\,,
\end{align}
and hence to reconstruct $\Delta E$ with increasing precision.

\section{Regulator choice}\label{app:reg}

The present work uses a simple flat or Litim regulator \cite{Litim:2001up} which is given by
\begin{align}\label{eq:reg}
	R_{k} = \left(k^2 - p^2 \right) \Theta\left(1-\frac{p^2}{k^2}\right) \,.
\end{align}
This regulator choice turns the evaluation of momentum loops analytical. In future works we envisage using smooth variants of the Litim regulator \cite{Ihssen:2024miv}. These functions are better suited to evaluate momentum dependent approximation schemes \cite{Pawlowski:2005xe}.

The standard Wetterich approach sometimes uses a modified regulator function $R_k \to Z_\varphi(\varphi_0) R_k$, where $Z_\varphi(\varphi_0)$ is the wave function evaluated on the equations of motion. 
By using the ground state expansion we have set $Z_\phi[\phi]=1$, which makes this modification unnecessary. Moreover, the PIRG approach regulates the composite field $ \phi$ thus effectively arranging for
\begin{align}
	\frac12 \int_p  \phi[ \varphi] R_k(p^2)  \phi[ \varphi] \approx \frac12 \int_p \varphi Z_\varphi(\varphi) R_k(p^2)  \varphi \,,
\end{align}
in the ground state approach with \labelcref{eq:Zphi=1}, without destroying the one-loop exactness of the Wetterich equation.

\section{Comparison to Wetterich flows}
\label{app:Wetterich1stod}
	
	In the present Section we outline the technical simplification of the ground state expansion \cite{Baldazzi:2021orb, Baldazzi:2021ydj, Ihssen:2023nqd} within the PIRG setup \cite{Ihssen:2024ihp}, in comparison to the baseline Wetterich flow. The Wetterich flow is obtained from \labelcref{eq:GenFlow} by setting the emergent composite operator to $\hat \phi = \hat \varphi$ and consequently $\dot \phi = 0$. For a clear distinction with the ground state expansion setup, we only use the notation in terms of the fundamental (mean) field $\varphi =\langle \hat \varphi\rangle$ in the following.
	
	We begin in \Cref{sec:TruncWett} by detailing the ansatz for the vertices within the first order derivative expansion of the standard Wetterich setup. This is followed by a discussion of the flows in \Cref{sec:WettDiags}.
	
\subsection{Truncation}\label{sec:TruncWett}
	
	The 1PI correlation functions can be deduced from the ansatz \labelcref{eq:Gamma1stOrder}, by evaluating the corresponding functional derivatives at constant fields $\varphi(\tau) = \varphi$. For example, the dispersion in frequency space is given by
	\begin{align}
		\Gamma_{\varphi}^{(2)}(\omega,-\omega)[\varphi] = \omega^2 Z_{\varphi} + \partial_\varphi^2 V_\varphi \,,
	\end{align}
	where the superscript $^{(n)}$ denotes the $n$th derivative with respect to the scalar field. For non-trivial $Z_\varphi$, this dispersion relation is not a classical one, since the frequency term has a corrective, field-dependent factor. This is due to the fact that $\varphi$ is not the renormalised physical field, which is needed for a simple description in terms of the ground state of the theory.
	
	The difficulty incurred by using the full first order derivative expansion becomes apparent at the example of the three- and four-point vertex functions
	\begin{subequations}\label{eq:highervertices}
		\begin{align}
			\Gamma_\varphi^{(3)}(\omega,\nu,-\omega-\nu)[\varphi] = (\omega^2 + \nu^2+ \omega \nu) \ \partial_\varphi Z_\varphi + \partial_\varphi^3 V_\varphi \,,
		\end{align}
		and 
		\begin{align}
			\Gamma_\varphi^{(4)}(\omega,-\omega,\nu,-\nu) = (\omega^2+ \nu^2 )\ \partial_\varphi^2 Z_\varphi + \partial_\varphi^4 V_\varphi \,.
		\end{align}
	\end{subequations}
	Additionally to the frequency independent scatterings contained in $V_k$-derivatives, \labelcref{eq:highervertices} contains frequency dependent contributions which are generated by the field-dependent wave function. These terms drop in the commonly used LPA' approximation scheme.
	
	Similarly, we have seen in \labelcref{eq:classDisp} that the higher vertices in the ground state expansion also do not have a momentum dependence. However, they correspond to the full first order of the derivative expansion, while maintaining the benefits of LPA'.

\subsection{Flow equations}
\label{sec:WettDiags}
	
Flow equations in the derivative expansion are evaluated by projecting on the corresponding momentum structure and evaluating again at constant fields. Within the first order derivative expansion and our present choice of regulator function, \Cref{app:reg}, the flow of the potential reads
\begin{subequations}\label{eq:wettflows}
		\begin{align}\label{eq:dtV} 
			\partial_t V_\varphi = 
			\frac{A_d \,k^{d+2}}{k^2 + V_\varphi^{(2)}} \, 
			\, _2F^1\left(1,d/2,1+d/2,\frac{k^2(1-Z_\varphi)}{k^2 + V_\varphi^{(2)}}\right)
			\,,
		\end{align}
		where $A_d$ is given by \labelcref{eq:Ad} and stems from the angular integration. $_2F^1$ is the hypergeometric function. For integer values of $d$ it reduces to expressions containing the $\tanh$- or $\log$-function. 
		
		The structure of \labelcref{eq:wettflows} can be simplified by choosing a regulator which contains the wave function as a prefactor. However, introducing field dependences to the regulator function turns the flow equation inexact and destroys the one-loop exact structure of the Wetterich equation.
		
		The flow of the wave function is obtained by taking a $\omega^2$ (frequency) derivative of the flow of the two-point function, and is given by
		\begin{align}\label{eq:dtZ}\notag 
			&\partial_t Z_\varphi(\varphi)
			= \frac{1}{2} \left. \partial_{\omega^2} \frac{\delta^2 \partial_t \Gamma_k[\varphi]}{\delta \varphi(-\omega) \delta \varphi(\omega)} \right|_{\omega=0} \notag \\[1ex]
			&=\int \frac{d \nu}{2\pi} \partial_t R_k(\nu) G^2_\varphi(\nu) \times \notag
			\frac{\partial}{\partial \omega^2} \Big\{ - \frac12 (\omega+\nu)^2 Z_\varphi^{(2)}\\[1ex]
			& \left.
			+\left[(\omega^2+\nu^2+\omega \nu) Z_\varphi^{(1)} + V_\varphi^{(3)}\right]^2 G_\varphi(\omega+\nu) 
			\Big\} \right|_{\omega=0}\,.
		\end{align}
	\end{subequations}
	The evaluation of the frequency derivative and the integration of the loop lead to additional expressions in terms of the hypergeometric function $_2 F^1$.
	Since the present work aims for an implementation in terms of the ground state expansion we refrain from solving this equation any further. For a pedagogic derivation of these flows, also including diagrams, see e.g.~\cite{Rennecke:2022ohx, Bonanno:2022edf}.

	\bibliographystyle{apsrev4-2}
	\bibliography{references}

\begin{thebibliography}{42}%
\makeatletter
\providecommand \@ifxundefined [1]{%
 \@ifx{#1\undefined}
}%
\providecommand \@ifnum [1]{%
 \ifnum #1\expandafter \@firstoftwo
 \else \expandafter \@secondoftwo
 \fi
}%
\providecommand \@ifx [1]{%
 \ifx #1\expandafter \@firstoftwo
 \else \expandafter \@secondoftwo
 \fi
}%
\providecommand \natexlab [1]{#1}%
\providecommand \enquote  [1]{``#1''}%
\providecommand \bibnamefont  [1]{#1}%
\providecommand \bibfnamefont [1]{#1}%
\providecommand \citenamefont [1]{#1}%
\providecommand \href@noop [0]{\@secondoftwo}%
\providecommand \href [0]{\begingroup \@sanitize@url \@href}%
\providecommand \@href[1]{\@@startlink{#1}\@@href}%
\providecommand \@@href[1]{\endgroup#1\@@endlink}%
\providecommand \@sanitize@url [0]{\catcode `\\12\catcode `\$12\catcode
  `\&12\catcode `\#12\catcode `\^12\catcode `\_12\catcode `\%12\relax}%
\providecommand \@@startlink[1]{}%
\providecommand \@@endlink[0]{}%
\providecommand \url  [0]{\begingroup\@sanitize@url \@url }%
\providecommand \@url [1]{\endgroup\@href {#1}{\urlprefix }}%
\providecommand \urlprefix  [0]{URL }%
\providecommand \Eprint [0]{\href }%
\providecommand \doibase [0]{https://doi.org/}%
\providecommand \selectlanguage [0]{\@gobble}%
\providecommand \bibinfo  [0]{\@secondoftwo}%
\providecommand \bibfield  [0]{\@secondoftwo}%
\providecommand \translation [1]{[#1]}%
\providecommand \BibitemOpen [0]{}%
\providecommand \bibitemStop [0]{}%
\providecommand \bibitemNoStop [0]{.\EOS\space}%
\providecommand \EOS [0]{\spacefactor3000\relax}%
\providecommand \BibitemShut  [1]{\csname bibitem#1\endcsname}%
\let\auto@bib@innerbib\@empty
\bibitem [{\citenamefont {Dupuis}\ \emph {et~al.}(2021)\citenamefont {Dupuis},
  \citenamefont {Canet}, \citenamefont {Eichhorn}, \citenamefont {Metzner},
  \citenamefont {Pawlowski}, \citenamefont {Tissier},\ and\ \citenamefont
  {Wschebor}}]{Dupuis:2020fhh}%
  \BibitemOpen
  \bibfield  {author} {\bibinfo {author} {\bibfnamefont {N.}~\bibnamefont
  {Dupuis}}, \bibinfo {author} {\bibfnamefont {L.}~\bibnamefont {Canet}},
  \bibinfo {author} {\bibfnamefont {A.}~\bibnamefont {Eichhorn}}, \bibinfo
  {author} {\bibfnamefont {W.}~\bibnamefont {Metzner}}, \bibinfo {author}
  {\bibfnamefont {J.~M.}\ \bibnamefont {Pawlowski}}, \bibinfo {author}
  {\bibfnamefont {M.}~\bibnamefont {Tissier}},\ and\ \bibinfo {author}
  {\bibfnamefont {N.}~\bibnamefont {Wschebor}},\ }\href
  {https://doi.org/10.1016/j.physrep.2021.01.001} {\bibfield  {journal}
  {\bibinfo  {journal} {Phys. Rept.}\ }\textbf {\bibinfo {volume} {910}},\
  \bibinfo {pages} {1} (\bibinfo {year} {2021})},\ \Eprint
  {https://arxiv.org/abs/2006.04853} {arXiv:2006.04853 [cond-mat.stat-mech]}
  \BibitemShut {NoStop}%
\bibitem [{\citenamefont {Zappala'}(2001)}]{Zappala:2001nv}%
  \BibitemOpen
  \bibfield  {author} {\bibinfo {author} {\bibfnamefont {D.}~\bibnamefont
  {Zappala'}},\ }\href {https://doi.org/10.1016/S0375-9601(01)00642-9}
  {\bibfield  {journal} {\bibinfo  {journal} {Phys. Lett. A}\ }\textbf
  {\bibinfo {volume} {290}},\ \bibinfo {pages} {35} (\bibinfo {year} {2001})},\
  \Eprint {https://arxiv.org/abs/quant-ph/0108019} {arXiv:quant-ph/0108019}
  \BibitemShut {NoStop}%
\bibitem [{\citenamefont {Aoki}\ \emph {et~al.}(2002)\citenamefont {Aoki},
  \citenamefont {Horikoshi}, \citenamefont {Taniguchi},\ and\ \citenamefont
  {Terao}}]{Aoki:2002ozs}%
  \BibitemOpen
  \bibfield  {author} {\bibinfo {author} {\bibfnamefont {K.-I.}\ \bibnamefont
  {Aoki}}, \bibinfo {author} {\bibfnamefont {A.}~\bibnamefont {Horikoshi}},
  \bibinfo {author} {\bibfnamefont {M.}~\bibnamefont {Taniguchi}},\ and\
  \bibinfo {author} {\bibfnamefont {H.}~\bibnamefont {Terao}},\ }\href
  {https://doi.org/10.1143/PTP.108.571} {\bibfield  {journal} {\bibinfo
  {journal} {Prog. Theor. Phys.}\ }\textbf {\bibinfo {volume} {108}},\ \bibinfo
  {pages} {571} (\bibinfo {year} {2002})},\ \Eprint
  {https://arxiv.org/abs/quant-ph/0208173} {arXiv:quant-ph/0208173}
  \BibitemShut {NoStop}%
\bibitem [{\citenamefont {Weyrauch}(2006)}]{Weyrauch:2006aj}%
  \BibitemOpen
  \bibfield  {author} {\bibinfo {author} {\bibfnamefont {M.}~\bibnamefont
  {Weyrauch}},\ }\href {https://doi.org/10.1088/0305-4470/39/3/015} {\bibfield
  {journal} {\bibinfo  {journal} {J. Phys. A}\ }\textbf {\bibinfo {volume}
  {39}},\ \bibinfo {pages} {649} (\bibinfo {year} {2006})}\BibitemShut
  {NoStop}%
\bibitem [{\citenamefont {Kovacs}\ \emph {et~al.}(2015)\citenamefont {Kovacs},
  \citenamefont {Nagy},\ and\ \citenamefont {Sailer}}]{Kovacs:2014mia}%
  \BibitemOpen
  \bibfield  {author} {\bibinfo {author} {\bibfnamefont {J.}~\bibnamefont
  {Kovacs}}, \bibinfo {author} {\bibfnamefont {S.}~\bibnamefont {Nagy}},\ and\
  \bibinfo {author} {\bibfnamefont {K.}~\bibnamefont {Sailer}},\ }\href
  {https://doi.org/10.1142/S0217751X1550058X} {\bibfield  {journal} {\bibinfo
  {journal} {Int. J. Mod. Phys. A}\ }\textbf {\bibinfo {volume} {30}},\
  \bibinfo {pages} {1550058} (\bibinfo {year} {2015})},\ \Eprint
  {https://arxiv.org/abs/1403.3544} {arXiv:1403.3544 [hep-th]} \BibitemShut
  {NoStop}%
\bibitem [{\citenamefont {Baldazzi}\ \emph {et~al.}(2021)\citenamefont
  {Baldazzi}, \citenamefont {Percacci},\ and\ \citenamefont
  {Zambelli}}]{Baldazzi:2021guw}%
  \BibitemOpen
  \bibfield  {author} {\bibinfo {author} {\bibfnamefont {A.}~\bibnamefont
  {Baldazzi}}, \bibinfo {author} {\bibfnamefont {R.}~\bibnamefont {Percacci}},\
  and\ \bibinfo {author} {\bibfnamefont {L.}~\bibnamefont {Zambelli}},\ }\href
  {https://doi.org/10.1103/PhysRevD.104.076026} {\bibfield  {journal} {\bibinfo
   {journal} {Phys. Rev. D}\ }\textbf {\bibinfo {volume} {104}},\ \bibinfo
  {pages} {076026} (\bibinfo {year} {2021})},\ \Eprint
  {https://arxiv.org/abs/2105.05778} {arXiv:2105.05778 [hep-th]} \BibitemShut
  {NoStop}%
\bibitem [{\citenamefont {Bonanno}\ \emph {et~al.}(2022)\citenamefont
  {Bonanno}, \citenamefont {Codello},\ and\ \citenamefont
  {Zappala'}}]{Bonanno:2022edf}%
  \BibitemOpen
  \bibfield  {author} {\bibinfo {author} {\bibfnamefont {A.}~\bibnamefont
  {Bonanno}}, \bibinfo {author} {\bibfnamefont {A.}~\bibnamefont {Codello}},\
  and\ \bibinfo {author} {\bibfnamefont {D.}~\bibnamefont {Zappala'}},\ }\href
  {https://doi.org/10.1016/j.aop.2022.169090} {\bibfield  {journal} {\bibinfo
  {journal} {Annals Phys.}\ }\textbf {\bibinfo {volume} {445}},\ \bibinfo
  {pages} {169090} (\bibinfo {year} {2022})},\ \Eprint
  {https://arxiv.org/abs/2206.06917} {arXiv:2206.06917 [hep-th]} \BibitemShut
  {NoStop}%
\bibitem [{\citenamefont {Ihssen}\ and\ \citenamefont
  {Pawlowski}(2024)}]{Ihssen:2024ihp}%
  \BibitemOpen
  \bibfield  {author} {\bibinfo {author} {\bibfnamefont {F.}~\bibnamefont
  {Ihssen}}\ and\ \bibinfo {author} {\bibfnamefont {J.~M.}\ \bibnamefont
  {Pawlowski}},\ }\href@noop {} {\  (\bibinfo {year} {2024})},\ \Eprint
  {https://arxiv.org/abs/2409.13679} {arXiv:2409.13679 [hep-th]} \BibitemShut
  {NoStop}%
\bibitem [{\citenamefont {Ihssen}\ and\ \citenamefont
  {Pawlowski}(2023)}]{Ihssen:2023nqd}%
  \BibitemOpen
  \bibfield  {author} {\bibinfo {author} {\bibfnamefont {F.}~\bibnamefont
  {Ihssen}}\ and\ \bibinfo {author} {\bibfnamefont {J.~M.}\ \bibnamefont
  {Pawlowski}},\ }\href@noop {} {\  (\bibinfo {year} {2023})},\ \Eprint
  {https://arxiv.org/abs/2305.00816} {arXiv:2305.00816 [hep-th]} \BibitemShut
  {NoStop}%
\bibitem [{\citenamefont {Ihssen}\ \emph
  {et~al.}(2024{\natexlab{a}})\citenamefont {Ihssen}, \citenamefont
  {Pawlowski}, \citenamefont {Sattler},\ and\ \citenamefont
  {Wink}}]{Ihssen:2022xkr}%
  \BibitemOpen
  \bibfield  {author} {\bibinfo {author} {\bibfnamefont {F.}~\bibnamefont
  {Ihssen}}, \bibinfo {author} {\bibfnamefont {J.~M.}\ \bibnamefont
  {Pawlowski}}, \bibinfo {author} {\bibfnamefont {F.~R.}\ \bibnamefont
  {Sattler}},\ and\ \bibinfo {author} {\bibfnamefont {N.}~\bibnamefont
  {Wink}},\ }\href {https://doi.org/10.1016/j.cpc.2024.109182} {\bibfield
  {journal} {\bibinfo  {journal} {Comput. Phys. Commun.}\ }\textbf {\bibinfo
  {volume} {300}},\ \bibinfo {pages} {109182} (\bibinfo {year}
  {2024}{\natexlab{a}})},\ \Eprint {https://arxiv.org/abs/2207.12266}
  {arXiv:2207.12266 [hep-th]} \BibitemShut {NoStop}%
\bibitem [{\citenamefont {Ihssen}\ \emph {et~al.}(2023)\citenamefont {Ihssen},
  \citenamefont {Sattler},\ and\ \citenamefont {Wink}}]{Ihssen:2023qaq}%
  \BibitemOpen
  \bibfield  {author} {\bibinfo {author} {\bibfnamefont {F.}~\bibnamefont
  {Ihssen}}, \bibinfo {author} {\bibfnamefont {F.~R.}\ \bibnamefont
  {Sattler}},\ and\ \bibinfo {author} {\bibfnamefont {N.}~\bibnamefont
  {Wink}},\ }\href {https://doi.org/10.1103/PhysRevD.107.114009} {\bibfield
  {journal} {\bibinfo  {journal} {Phys. Rev. D}\ }\textbf {\bibinfo {volume}
  {107}},\ \bibinfo {pages} {114009} (\bibinfo {year} {2023})},\ \Eprint
  {https://arxiv.org/abs/2302.04736} {arXiv:2302.04736 [hep-th]} \BibitemShut
  {NoStop}%
\bibitem [{\citenamefont {Sattler}\ and\ \citenamefont
  {Pawlowski}(2024)}]{Sattler:2024ozv}%
  \BibitemOpen
  \bibfield  {author} {\bibinfo {author} {\bibfnamefont {F.~R.}\ \bibnamefont
  {Sattler}}\ and\ \bibinfo {author} {\bibfnamefont {J.~M.}\ \bibnamefont
  {Pawlowski}},\ }\href@noop {} {\  (\bibinfo {year} {2024})},\ \Eprint
  {https://arxiv.org/abs/2412.13043} {arXiv:2412.13043 [hep-ph]} \BibitemShut
  {NoStop}%
\bibitem [{\citenamefont {Sch\"afer}\ and\ \citenamefont
  {Shuryak}(1998)}]{Schafer:1996wv}%
  \BibitemOpen
  \bibfield  {author} {\bibinfo {author} {\bibfnamefont {T.}~\bibnamefont
  {Sch\"afer}}\ and\ \bibinfo {author} {\bibfnamefont {E.~V.}\ \bibnamefont
  {Shuryak}},\ }\href {https://doi.org/10.1103/RevModPhys.70.323} {\bibfield
  {journal} {\bibinfo  {journal} {Rev. Mod. Phys.}\ }\textbf {\bibinfo {volume}
  {70}},\ \bibinfo {pages} {323} (\bibinfo {year} {1998})},\ \Eprint
  {https://arxiv.org/abs/hep-ph/9610451} {arXiv:hep-ph/9610451} \BibitemShut
  {NoStop}%
\bibitem [{\citenamefont {Pawlowski}(2007)}]{Pawlowski:2005xe}%
  \BibitemOpen
  \bibfield  {author} {\bibinfo {author} {\bibfnamefont {J.~M.}\ \bibnamefont
  {Pawlowski}},\ }\href {https://doi.org/10.1016/j.aop.2007.01.007} {\bibfield
  {journal} {\bibinfo  {journal} {Annals Phys.}\ }\textbf {\bibinfo {volume}
  {322}},\ \bibinfo {pages} {2831} (\bibinfo {year} {2007})},\ \Eprint
  {https://arxiv.org/abs/hep-th/0512261} {arXiv:hep-th/0512261} \BibitemShut
  {NoStop}%
\bibitem [{\citenamefont {Wegner}(1974)}]{Wegner_1974}%
  \BibitemOpen
  \bibfield  {author} {\bibinfo {author} {\bibfnamefont {F.~J.}\ \bibnamefont
  {Wegner}},\ }\href {https://doi.org/10.1088/0022-3719/7/12/004} {\bibfield
  {journal} {\bibinfo  {journal} {Journal of Physics C: Solid State Physics}\
  }\textbf {\bibinfo {volume} {7}},\ \bibinfo {pages} {2098} (\bibinfo {year}
  {1974})}\BibitemShut {NoStop}%
\bibitem [{\citenamefont {Polchinski}(1984)}]{Polchinski1984}%
  \BibitemOpen
  \bibfield  {author} {\bibinfo {author} {\bibfnamefont {J.}~\bibnamefont
  {Polchinski}},\ }\href {https://doi.org/10.1016/0550-3213(84)90287-6}
  {\bibfield  {journal} {\bibinfo  {journal} {Nuclear Physics B}\ }\textbf
  {\bibinfo {volume} {231}},\ \bibinfo {pages} {269} (\bibinfo {year}
  {1984})}\BibitemShut {NoStop}%
\bibitem [{\citenamefont {Wetterich}(1993)}]{Wetterich:1992yh}%
  \BibitemOpen
  \bibfield  {author} {\bibinfo {author} {\bibfnamefont {C.}~\bibnamefont
  {Wetterich}},\ }\href {https://doi.org/10.1016/0370-2693(93)90726-X}
  {\bibfield  {journal} {\bibinfo  {journal} {Phys. Lett. B}\ }\textbf
  {\bibinfo {volume} {301}},\ \bibinfo {pages} {90} (\bibinfo {year} {1993})},\
  \Eprint {https://arxiv.org/abs/1710.05815} {arXiv:1710.05815 [hep-th]}
  \BibitemShut {NoStop}%
\bibitem [{\citenamefont {Ihssen}\ and\ \citenamefont
  {Pawlowski}(2025)}]{Ihssen:2025cff}%
  \BibitemOpen
  \bibfield  {author} {\bibinfo {author} {\bibfnamefont {F.}~\bibnamefont
  {Ihssen}}\ and\ \bibinfo {author} {\bibfnamefont {J.~M.}\ \bibnamefont
  {Pawlowski}},\ }\href@noop {} {\  (\bibinfo {year} {2025})},\ \Eprint
  {https://arxiv.org/abs/2503.22638} {arXiv:2503.22638 [hep-th]} \BibitemShut
  {NoStop}%
\bibitem [{\citenamefont {Ihssen}\ \emph {et~al.}(2025)\citenamefont {Ihssen},
  \citenamefont {Pawlowski}, \citenamefont {Sattler},\ and\ \citenamefont
  {Wink}}]{Ihssen:2023xlp}%
  \BibitemOpen
  \bibfield  {author} {\bibinfo {author} {\bibfnamefont {F.}~\bibnamefont
  {Ihssen}}, \bibinfo {author} {\bibfnamefont {J.~M.}\ \bibnamefont
  {Pawlowski}}, \bibinfo {author} {\bibfnamefont {F.~R.}\ \bibnamefont
  {Sattler}},\ and\ \bibinfo {author} {\bibfnamefont {N.}~\bibnamefont
  {Wink}},\ }\href {https://doi.org/10.1103/PhysRevD.111.036030} {\bibfield
  {journal} {\bibinfo  {journal} {Phys. Rev. D}\ }\textbf {\bibinfo {volume}
  {111}},\ \bibinfo {pages} {036030} (\bibinfo {year} {2025})},\ \Eprint
  {https://arxiv.org/abs/2309.07335} {arXiv:2309.07335 [hep-th]} \BibitemShut
  {NoStop}%
\bibitem [{\citenamefont {Grossi}\ and\ \citenamefont
  {Wink}(2023)}]{Grossi:2019urj}%
  \BibitemOpen
  \bibfield  {author} {\bibinfo {author} {\bibfnamefont {E.}~\bibnamefont
  {Grossi}}\ and\ \bibinfo {author} {\bibfnamefont {N.}~\bibnamefont {Wink}},\
  }\href {https://doi.org/10.21468/SciPostPhysCore.6.4.071} {\bibfield
  {journal} {\bibinfo  {journal} {SciPost Phys. Core}\ }\textbf {\bibinfo
  {volume} {6}},\ \bibinfo {pages} {071} (\bibinfo {year} {2023})},\ \Eprint
  {https://arxiv.org/abs/1903.09503} {arXiv:1903.09503 [hep-th]} \BibitemShut
  {NoStop}%
\bibitem [{\citenamefont {Grossi}\ \emph {et~al.}(2021)\citenamefont {Grossi},
  \citenamefont {Ihssen}, \citenamefont {Pawlowski},\ and\ \citenamefont
  {Wink}}]{Grossi:2021ksl}%
  \BibitemOpen
  \bibfield  {author} {\bibinfo {author} {\bibfnamefont {E.}~\bibnamefont
  {Grossi}}, \bibinfo {author} {\bibfnamefont {F.~J.}\ \bibnamefont {Ihssen}},
  \bibinfo {author} {\bibfnamefont {J.~M.}\ \bibnamefont {Pawlowski}},\ and\
  \bibinfo {author} {\bibfnamefont {N.}~\bibnamefont {Wink}},\ }\href
  {https://doi.org/10.1103/PhysRevD.104.016028} {\bibfield  {journal} {\bibinfo
   {journal} {Phys. Rev. D}\ }\textbf {\bibinfo {volume} {104}},\ \bibinfo
  {pages} {016028} (\bibinfo {year} {2021})},\ \Eprint
  {https://arxiv.org/abs/2102.01602} {arXiv:2102.01602 [hep-ph]} \BibitemShut
  {NoStop}%
\bibitem [{\citenamefont {Koenigstein}\ \emph
  {et~al.}(2022{\natexlab{a}})\citenamefont {Koenigstein}, \citenamefont
  {Steil}, \citenamefont {Wink}, \citenamefont {Grossi},\ and\ \citenamefont
  {Braun}}]{Koenigstein:2021rxj}%
  \BibitemOpen
  \bibfield  {author} {\bibinfo {author} {\bibfnamefont {A.}~\bibnamefont
  {Koenigstein}}, \bibinfo {author} {\bibfnamefont {M.~J.}\ \bibnamefont
  {Steil}}, \bibinfo {author} {\bibfnamefont {N.}~\bibnamefont {Wink}},
  \bibinfo {author} {\bibfnamefont {E.}~\bibnamefont {Grossi}},\ and\ \bibinfo
  {author} {\bibfnamefont {J.}~\bibnamefont {Braun}},\ }\href
  {https://doi.org/10.1103/PhysRevD.106.065013} {\bibfield  {journal} {\bibinfo
   {journal} {Phys. Rev. D}\ }\textbf {\bibinfo {volume} {106}},\ \bibinfo
  {pages} {065013} (\bibinfo {year} {2022}{\natexlab{a}})},\ \Eprint
  {https://arxiv.org/abs/2108.10085} {arXiv:2108.10085 [cond-mat.stat-mech]}
  \BibitemShut {NoStop}%
\bibitem [{\citenamefont {Koenigstein}\ \emph
  {et~al.}(2022{\natexlab{b}})\citenamefont {Koenigstein}, \citenamefont
  {Steil}, \citenamefont {Wink}, \citenamefont {Grossi}, \citenamefont {Braun},
  \citenamefont {Buballa},\ and\ \citenamefont
  {Rischke}}]{Koenigstein:2021syz}%
  \BibitemOpen
  \bibfield  {author} {\bibinfo {author} {\bibfnamefont {A.}~\bibnamefont
  {Koenigstein}}, \bibinfo {author} {\bibfnamefont {M.~J.}\ \bibnamefont
  {Steil}}, \bibinfo {author} {\bibfnamefont {N.}~\bibnamefont {Wink}},
  \bibinfo {author} {\bibfnamefont {E.}~\bibnamefont {Grossi}}, \bibinfo
  {author} {\bibfnamefont {J.}~\bibnamefont {Braun}}, \bibinfo {author}
  {\bibfnamefont {M.}~\bibnamefont {Buballa}},\ and\ \bibinfo {author}
  {\bibfnamefont {D.~H.}\ \bibnamefont {Rischke}},\ }\href
  {https://doi.org/10.1103/PhysRevD.106.065012} {\bibfield  {journal} {\bibinfo
   {journal} {Phys. Rev. D}\ }\textbf {\bibinfo {volume} {106}},\ \bibinfo
  {pages} {065012} (\bibinfo {year} {2022}{\natexlab{b}})},\ \Eprint
  {https://arxiv.org/abs/2108.02504} {arXiv:2108.02504 [cond-mat.stat-mech]}
  \BibitemShut {NoStop}%
\bibitem [{\citenamefont {Steil}\ and\ \citenamefont
  {Koenigstein}(2022)}]{Steil:2021cbu}%
  \BibitemOpen
  \bibfield  {author} {\bibinfo {author} {\bibfnamefont {M.~J.}\ \bibnamefont
  {Steil}}\ and\ \bibinfo {author} {\bibfnamefont {A.}~\bibnamefont
  {Koenigstein}},\ }\href {https://doi.org/10.1103/PhysRevD.106.065014}
  {\bibfield  {journal} {\bibinfo  {journal} {Phys. Rev. D}\ }\textbf {\bibinfo
  {volume} {106}},\ \bibinfo {pages} {065014} (\bibinfo {year} {2022})},\
  \Eprint {https://arxiv.org/abs/2108.04037} {arXiv:2108.04037
  [cond-mat.stat-mech]} \BibitemShut {NoStop}%
\bibitem [{\citenamefont {Koenigstein}\ and\ \citenamefont
  {Pannullo}(2024)}]{Koenigstein:2023yzv}%
  \BibitemOpen
  \bibfield  {author} {\bibinfo {author} {\bibfnamefont {A.}~\bibnamefont
  {Koenigstein}}\ and\ \bibinfo {author} {\bibfnamefont {L.}~\bibnamefont
  {Pannullo}},\ }\href {https://doi.org/10.1103/PhysRevD.109.056015} {\bibfield
   {journal} {\bibinfo  {journal} {Phys. Rev. D}\ }\textbf {\bibinfo {volume}
  {109}},\ \bibinfo {pages} {056015} (\bibinfo {year} {2024})},\ \Eprint
  {https://arxiv.org/abs/2312.04904} {arXiv:2312.04904 [hep-ph]} \BibitemShut
  {NoStop}%
\bibitem [{\citenamefont {Ihssen}\ \emph
  {et~al.}(2024{\natexlab{b}})\citenamefont {Ihssen}, \citenamefont
  {Pawlowski}, \citenamefont {Sattler},\ and\ \citenamefont
  {Wink}}]{Ihssen:2024miv}%
  \BibitemOpen
  \bibfield  {author} {\bibinfo {author} {\bibfnamefont {F.}~\bibnamefont
  {Ihssen}}, \bibinfo {author} {\bibfnamefont {J.~M.}\ \bibnamefont
  {Pawlowski}}, \bibinfo {author} {\bibfnamefont {F.~R.}\ \bibnamefont
  {Sattler}},\ and\ \bibinfo {author} {\bibfnamefont {N.}~\bibnamefont
  {Wink}},\ }\href@noop {} {\  (\bibinfo {year} {2024}{\natexlab{b}})},\
  \Eprint {https://arxiv.org/abs/2408.08413} {arXiv:2408.08413 [hep-ph]}
  \BibitemShut {NoStop}%
\bibitem [{\citenamefont {Zorbach}\ \emph {et~al.}(2024)\citenamefont
  {Zorbach}, \citenamefont {Koenigstein},\ and\ \citenamefont
  {Braun}}]{Zorbach:2024rre}%
  \BibitemOpen
  \bibfield  {author} {\bibinfo {author} {\bibfnamefont {N.}~\bibnamefont
  {Zorbach}}, \bibinfo {author} {\bibfnamefont {A.}~\bibnamefont
  {Koenigstein}},\ and\ \bibinfo {author} {\bibfnamefont {J.}~\bibnamefont
  {Braun}},\ }\href@noop {} {\  (\bibinfo {year} {2024})},\ \Eprint
  {https://arxiv.org/abs/2412.16053} {arXiv:2412.16053 [cond-mat.stat-mech]}
  \BibitemShut {NoStop}%
\bibitem [{\citenamefont {Berezinskii}(1971)}]{berezinskii1971destruction}%
  \BibitemOpen
  \bibfield  {author} {\bibinfo {author} {\bibfnamefont {V.~L.}\ \bibnamefont
  {Berezinskii}},\ }\href@noop {} {\bibfield  {journal} {\bibinfo  {journal}
  {Soviet Journal of Experimental and Theoretical Physics}\ }\textbf {\bibinfo
  {volume} {32}},\ \bibinfo {pages} {493} (\bibinfo {year} {1971})}\BibitemShut
  {NoStop}%
\bibitem [{\citenamefont {Berezinskii}(1972)}]{berezinskii1972destruction}%
  \BibitemOpen
  \bibfield  {author} {\bibinfo {author} {\bibfnamefont {V.~L.}\ \bibnamefont
  {Berezinskii}},\ }\href@noop {} {\bibfield  {journal} {\bibinfo  {journal}
  {Soviet Journal of Experimental and Theoretical Physics}\ }\textbf {\bibinfo
  {volume} {34}},\ \bibinfo {pages} {610} (\bibinfo {year} {1972})}\BibitemShut
  {NoStop}%
\bibitem [{\citenamefont {Kosterlitz}\ and\ \citenamefont
  {Thouless}(1973)}]{kosterlitz1973ordering}%
  \BibitemOpen
  \bibfield  {author} {\bibinfo {author} {\bibfnamefont {J.~M.}\ \bibnamefont
  {Kosterlitz}}\ and\ \bibinfo {author} {\bibfnamefont {D.~J.}\ \bibnamefont
  {Thouless}},\ }\href {https://doi.org/10.1088/0022-3719/6/7/010} {\bibfield
  {journal} {\bibinfo  {journal} {Journal of Physics C: Solid State Physics}\
  }\textbf {\bibinfo {volume} {6}},\ \bibinfo {pages} {1181} (\bibinfo {year}
  {1973})}\BibitemShut {NoStop}%
\bibitem [{\citenamefont {Kosterlitz}(1974)}]{kosterlitz1974critical}%
  \BibitemOpen
  \bibfield  {author} {\bibinfo {author} {\bibfnamefont {J.~M.}\ \bibnamefont
  {Kosterlitz}},\ }\href {https://doi.org/10.1088/0022-3719/7/6/005} {\bibfield
   {journal} {\bibinfo  {journal} {Journal of Physics C: Solid State Physics}\
  }\textbf {\bibinfo {volume} {7}},\ \bibinfo {pages} {1046} (\bibinfo {year}
  {1974})}\BibitemShut {NoStop}%
\bibitem [{\citenamefont {Grater}\ and\ \citenamefont
  {Wetterich}(1995)}]{Grater:1994qx}%
  \BibitemOpen
  \bibfield  {author} {\bibinfo {author} {\bibfnamefont {M.}~\bibnamefont
  {Grater}}\ and\ \bibinfo {author} {\bibfnamefont {C.}~\bibnamefont
  {Wetterich}},\ }\href {https://doi.org/10.1103/PhysRevLett.75.378} {\bibfield
   {journal} {\bibinfo  {journal} {Phys. Rev. Lett.}\ }\textbf {\bibinfo
  {volume} {75}},\ \bibinfo {pages} {378} (\bibinfo {year} {1995})},\ \Eprint
  {https://arxiv.org/abs/hep-ph/9409459} {arXiv:hep-ph/9409459} \BibitemShut
  {NoStop}%
\bibitem [{\citenamefont {Von~Gersdorff}\ and\ \citenamefont
  {Wetterich}(2001)}]{VonGersdorff:2000kp}%
  \BibitemOpen
  \bibfield  {author} {\bibinfo {author} {\bibfnamefont {G.}~\bibnamefont
  {Von~Gersdorff}}\ and\ \bibinfo {author} {\bibfnamefont {C.}~\bibnamefont
  {Wetterich}},\ }\href {https://doi.org/10.1103/PhysRevB.64.054513} {\bibfield
   {journal} {\bibinfo  {journal} {Phys. Rev. B}\ }\textbf {\bibinfo {volume}
  {64}},\ \bibinfo {pages} {054513} (\bibinfo {year} {2001})},\ \Eprint
  {https://arxiv.org/abs/hep-th/0008114} {arXiv:hep-th/0008114} \BibitemShut
  {NoStop}%
\bibitem [{\citenamefont {Jakubczyk}\ \emph {et~al.}(2014)\citenamefont
  {Jakubczyk}, \citenamefont {Dupuis},\ and\ \citenamefont
  {Delamotte}}]{Jakubczyk:2014isa}%
  \BibitemOpen
  \bibfield  {author} {\bibinfo {author} {\bibfnamefont {P.}~\bibnamefont
  {Jakubczyk}}, \bibinfo {author} {\bibfnamefont {N.}~\bibnamefont {Dupuis}},\
  and\ \bibinfo {author} {\bibfnamefont {B.}~\bibnamefont {Delamotte}},\ }\href
  {https://doi.org/10.1103/PhysRevE.90.062105} {\bibfield  {journal} {\bibinfo
  {journal} {Phys. Rev. E}\ }\textbf {\bibinfo {volume} {90}},\ \bibinfo
  {pages} {062105} (\bibinfo {year} {2014})},\ \Eprint
  {https://arxiv.org/abs/1409.1374} {arXiv:1409.1374 [cond-mat.stat-mech]}
  \BibitemShut {NoStop}%
\bibitem [{\citenamefont {Jakubczyk}\ and\ \citenamefont
  {Metzner}(2017)}]{Jakubczyk:2016rvr}%
  \BibitemOpen
  \bibfield  {author} {\bibinfo {author} {\bibfnamefont {P.}~\bibnamefont
  {Jakubczyk}}\ and\ \bibinfo {author} {\bibfnamefont {W.}~\bibnamefont
  {Metzner}},\ }\href {https://doi.org/10.1103/PhysRevB.95.085113} {\bibfield
  {journal} {\bibinfo  {journal} {Phys. Rev. B}\ }\textbf {\bibinfo {volume}
  {95}},\ \bibinfo {pages} {085113} (\bibinfo {year} {2017})},\ \Eprint
  {https://arxiv.org/abs/1606.04547} {arXiv:1606.04547 [cond-mat.stat-mech]}
  \BibitemShut {NoStop}%
\bibitem [{\citenamefont {Defenu}\ \emph {et~al.}(2017)\citenamefont {Defenu},
  \citenamefont {Trombettoni}, \citenamefont {N\'andori},\ and\ \citenamefont
  {Enss}}]{Defenu:2017}%
  \BibitemOpen
  \bibfield  {author} {\bibinfo {author} {\bibfnamefont {N.}~\bibnamefont
  {Defenu}}, \bibinfo {author} {\bibfnamefont {A.}~\bibnamefont {Trombettoni}},
  \bibinfo {author} {\bibfnamefont {I.}~\bibnamefont {N\'andori}},\ and\
  \bibinfo {author} {\bibfnamefont {T.}~\bibnamefont {Enss}},\ }\href
  {https://doi.org/10.1103/PhysRevB.96.174505} {\bibfield  {journal} {\bibinfo
  {journal} {Phys. Rev. B}\ }\textbf {\bibinfo {volume} {96}},\ \bibinfo
  {pages} {174505} (\bibinfo {year} {2017})}\BibitemShut {NoStop}%
\bibitem [{\citenamefont {Giachetti}\ \emph {et~al.}(2022)\citenamefont
  {Giachetti}, \citenamefont {Trombettoni}, \citenamefont {Ruffo},\ and\
  \citenamefont {Defenu}}]{Giachetti:2021woq}%
  \BibitemOpen
  \bibfield  {author} {\bibinfo {author} {\bibfnamefont {G.}~\bibnamefont
  {Giachetti}}, \bibinfo {author} {\bibfnamefont {A.}~\bibnamefont
  {Trombettoni}}, \bibinfo {author} {\bibfnamefont {S.}~\bibnamefont {Ruffo}},\
  and\ \bibinfo {author} {\bibfnamefont {N.}~\bibnamefont {Defenu}},\ }\href
  {https://doi.org/10.1103/PhysRevB.106.014106} {\bibfield  {journal} {\bibinfo
   {journal} {Phys. Rev. B}\ }\textbf {\bibinfo {volume} {106}},\ \bibinfo
  {pages} {014106} (\bibinfo {year} {2022})},\ \Eprint
  {https://arxiv.org/abs/2201.03650} {arXiv:2201.03650 [cond-mat.stat-mech]}
  \BibitemShut {NoStop}%
\bibitem [{\citenamefont {Bonanno}\ \emph {et~al.}(2020)\citenamefont
  {Bonanno}, \citenamefont {Lippoldt}, \citenamefont {Percacci},\ and\
  \citenamefont {Vacca}}]{Bonanno:2019ukb}%
  \BibitemOpen
  \bibfield  {author} {\bibinfo {author} {\bibfnamefont {A.}~\bibnamefont
  {Bonanno}}, \bibinfo {author} {\bibfnamefont {S.}~\bibnamefont {Lippoldt}},
  \bibinfo {author} {\bibfnamefont {R.}~\bibnamefont {Percacci}},\ and\
  \bibinfo {author} {\bibfnamefont {G.~P.}\ \bibnamefont {Vacca}},\ }\href
  {https://doi.org/10.1140/epjc/s10052-020-7798-9} {\bibfield  {journal}
  {\bibinfo  {journal} {Eur. Phys. J. C}\ }\textbf {\bibinfo {volume} {80}},\
  \bibinfo {pages} {249} (\bibinfo {year} {2020})},\ \Eprint
  {https://arxiv.org/abs/1912.08135} {arXiv:1912.08135 [hep-th]} \BibitemShut
  {NoStop}%
\bibitem [{\citenamefont {Litim}(2001)}]{Litim:2001up}%
  \BibitemOpen
  \bibfield  {author} {\bibinfo {author} {\bibfnamefont {D.~F.}\ \bibnamefont
  {Litim}},\ }\href {https://doi.org/10.1103/PhysRevD.64.105007} {\bibfield
  {journal} {\bibinfo  {journal} {Phys. Rev. D}\ }\textbf {\bibinfo {volume}
  {64}},\ \bibinfo {pages} {105007} (\bibinfo {year} {2001})},\ \Eprint
  {https://arxiv.org/abs/hep-th/0103195} {arXiv:hep-th/0103195} \BibitemShut
  {NoStop}%
\bibitem [{\citenamefont {Baldazzi}\ and\ \citenamefont
  {Falls}(2021)}]{Baldazzi:2021orb}%
  \BibitemOpen
  \bibfield  {author} {\bibinfo {author} {\bibfnamefont {A.}~\bibnamefont
  {Baldazzi}}\ and\ \bibinfo {author} {\bibfnamefont {K.}~\bibnamefont
  {Falls}},\ }\href {https://doi.org/10.3390/universe7080294} {\bibfield
  {journal} {\bibinfo  {journal} {Universe}\ }\textbf {\bibinfo {volume} {7}},\
  \bibinfo {pages} {294} (\bibinfo {year} {2021})},\ \Eprint
  {https://arxiv.org/abs/2107.00671} {arXiv:2107.00671 [hep-th]} \BibitemShut
  {NoStop}%
\bibitem [{\citenamefont {Baldazzi}\ \emph {et~al.}(2022)\citenamefont
  {Baldazzi}, \citenamefont {Zinati},\ and\ \citenamefont
  {Falls}}]{Baldazzi:2021ydj}%
  \BibitemOpen
  \bibfield  {author} {\bibinfo {author} {\bibfnamefont {A.}~\bibnamefont
  {Baldazzi}}, \bibinfo {author} {\bibfnamefont {R.~B.~A.}\ \bibnamefont
  {Zinati}},\ and\ \bibinfo {author} {\bibfnamefont {K.}~\bibnamefont
  {Falls}},\ }\href {https://doi.org/10.21468/SciPostPhys.13.4.085} {\bibfield
  {journal} {\bibinfo  {journal} {SciPost Phys.}\ }\textbf {\bibinfo {volume}
  {13}},\ \bibinfo {pages} {085} (\bibinfo {year} {2022})},\ \Eprint
  {https://arxiv.org/abs/2105.11482} {arXiv:2105.11482 [hep-th]} \BibitemShut
  {NoStop}%
\bibitem [{\citenamefont {Rennecke}\ and\ \citenamefont
  {Skokov}(2022)}]{Rennecke:2022ohx}%
  \BibitemOpen
  \bibfield  {author} {\bibinfo {author} {\bibfnamefont {F.}~\bibnamefont
  {Rennecke}}\ and\ \bibinfo {author} {\bibfnamefont {V.~V.}\ \bibnamefont
  {Skokov}},\ }\href {https://doi.org/10.1016/j.aop.2022.169010} {\bibfield
  {journal} {\bibinfo  {journal} {Annals Phys.}\ }\textbf {\bibinfo {volume}
  {444}},\ \bibinfo {pages} {169010} (\bibinfo {year} {2022})},\ \Eprint
  {https://arxiv.org/abs/2203.16651} {arXiv:2203.16651 [hep-ph]} \BibitemShut
  {NoStop}%
\end{thebibliography}%
	
\end{document}